\documentclass[apl,reprint,twocolumn]{revtex4}
\usepackage{graphicx}
\usepackage{amsmath}
\usepackage{amssymb}
\usepackage[english]{babel}
\usepackage{xcolor}
\usepackage{siunitx}

\usepackage[colorlinks=true, pdfstartview=FitV, linkcolor=blue,
citecolor=blue, urlcolor=blue]{hyperref}
\usepackage[capitalise]{cleveref}

\makeatletter
\let\saved@includegraphics\includegraphics
\AtBeginDocument{\let\includegraphics\saved@includegraphics}
\renewenvironment*{figure}{\@float{figure}}{\end@float}
\renewenvironment*{table}{\@float{table}}{\end@float}
\makeatother

\newcommand\T{\rule{0pt}{2.6ex}}       
\newcommand\B{\rule[-1.2ex]{0pt}{0pt}} 

\begin{document}
\title{Force sensing with nanowire cantilevers}
\author{F. R. Braakman$^{1}$}\thanks{Author to whom correspondence should be addressed.  Electronic mail: \texttt{floris.braakman@unibas.ch}.}
\author{M. Poggio$^{1}$}
\affiliation{1: University of Basel, Klingelbergstrasse 82, 4056 Basel, Switzerland}

\date{\today}

\begin{abstract}
\noindent Nanometer-scale structures with high aspect ratio such as nanowires and nanotubes combine low mechanical dissipation with high resonance frequencies, making them ideal force transducers and scanning probes in applications requiring the highest sensitivity. Such structures promise record force sensitivities combined with ease of use in scanning probe microscopes. A wide variety of possible material compositions and functionalizations is available, allowing for the sensing of various kinds of forces with optimized sensitivity. In addition, nanowires possess quasi-degenerate mechanical mode doublets, which has allowed the demonstration of sensitive vectorial force and mass detection. These developments have driven researchers to use nanowire cantilevers in various force sensing applications, which include imaging of sample surface topography, detection of optomechanical, electrical, and magnetic forces, and magnetic resonance force microscopy. In this review, we discuss the motivation behind using nanowires as force transducers, explain the methods of force sensing with nanowire cantilevers, and give an overview of the experimental progress and future prospects of the field.
\end{abstract}
\maketitle

\tableofcontents

\newpage

\section{Introduction}
\label{Sec:intro}
Recent years have seen a dramatic reduction in the size of mechanical
elements that can be used as mass and force sensors. For the most
sensitive applications, micro-processed Si cantilevers are starting to
give way to bottom-up fabricated structures such as nanowires
(NWs) and carbon nanotubes (CNTs). Bottom-up processes rely on
self-assembly or driven self-assembly and allow for the production of
nanometer-scale structures with atomic-scale precision. This trend
towards miniaturization is not arbitrary: smaller mechanical
transducers are inherently more sensitive. At the same time,
atomic-scale control in the growth of such structures presents the
opportunity to drastically reduce defects, improving mechanical
quality and therefore also ultimate detection sensitivity.  

Among the large variety of bottom-up nanostructures, NW cantilevers
are particularly promising mechanical sensors due to their high aspect
ratio.  A long and thin NW that is clamped on one end forms an ideal
scanning probe, making it amenable to a number of sensitive
nanometer-scale imaging problems.  At the same time, its symmetric
cross-section results in orthogonal flexural mode doublets that are
nearly degenerate. These modes allow a NW to be used as a kind of
nanometer-scale force compass~\cite{GloppeBidimensionalnanooptomechanicstopological2014}, such
that both the magnitude and direction of a force can be
measured. When the NW axis is oriented perpendicular to a sample surface, i.e. in the pendulum geometry, this enables the vectorial transduction of lateral forces. Moroever, the pendulum geometry allows a significant reduction of surface-induced dissipation compared to the more conventional parallel configuration. Furthermore, epitaxial growth allows the realization of NWs from a number of materials as well as heterostructures of such materials. In fact, due to their large surface-to-volume ratio, NWs
are able to accommodate larger strain than conventional epitaxial
films, allowing for the dislocation-free combination of materials with
large lattice mismatch~\cite{ThillosenStateStrainSingle2006}.

\begin{figure*}[t]
	\includegraphics[width=0.95\textwidth]{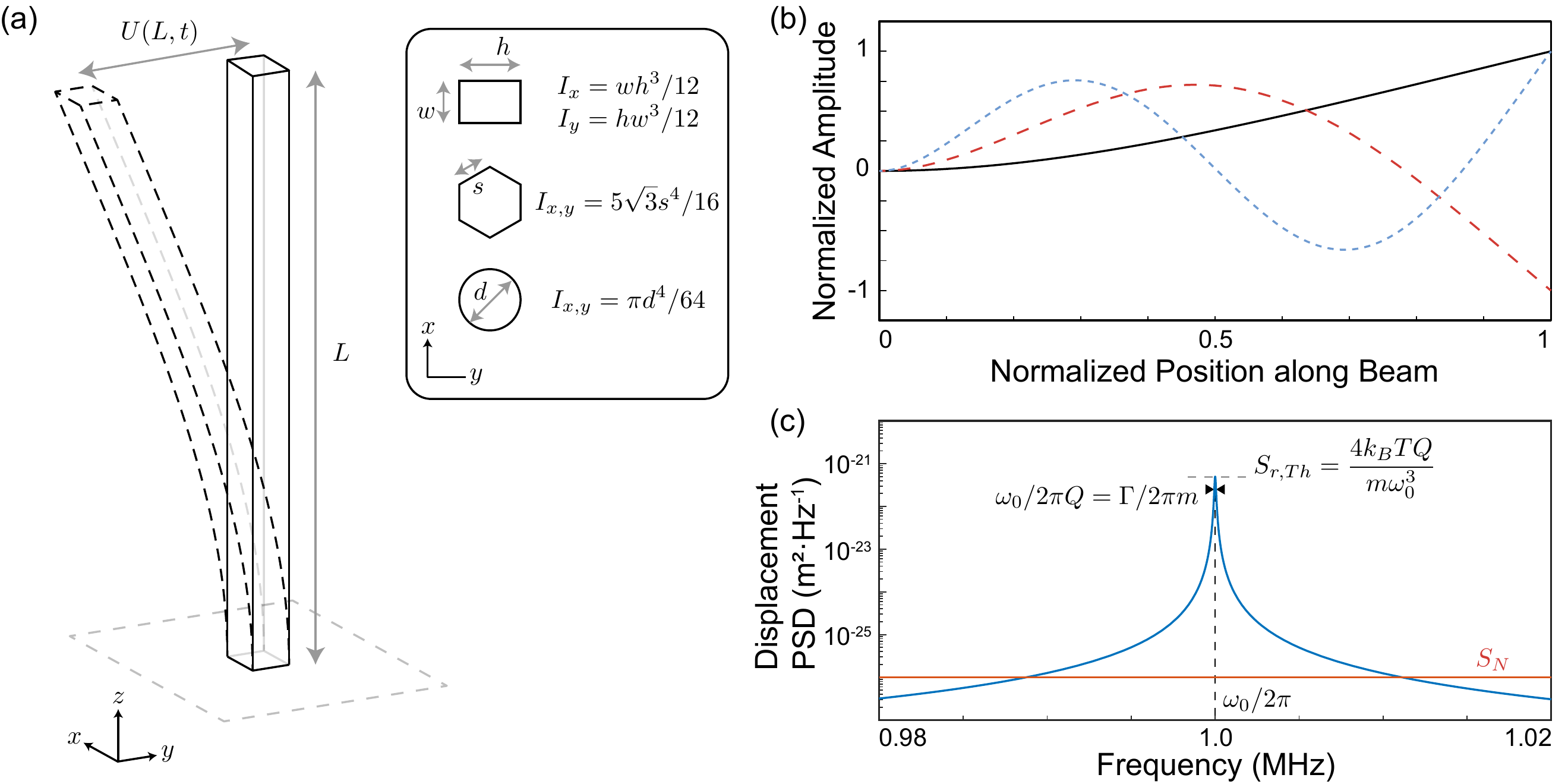}
	\caption{(a) Schematic overview of a singly clamped beam. Inset shows three cross-sectional beam shapes and corresponding moments of inertia. (b) Mode shapes of the first three flexural modes of a singly-clamped beam. (c) Plot of displacement power spectral density of a thermally driven resonator, for typical values of $Q$, $\omega_0$ and $S_{F,Th}$. $S_{N}$ is the power spectral density of measurement noise. Arrows indicate the half maximum.\vspace{1em}}
	\label{fig:mechanics}
\end{figure*}
The application of NWs to force sensing is proving particularly
apt. As highly sensitive scanning probes, NW sensors have the ability
to reveal subtle force fields with high spatial resolution. Unlike
conventional atomic force microscopy (AFM), which reveals large,
close-range forces -- in some cases with atomic resolution -- NWs are
adept at discerning weak interactions.  Ultra-low
intrinsic mechanical losses make NWs excellent probes of tip-sample
dissipation, a contrast which can be used to observe phase transitions
as well as the local density of
states~\cite{GrutterMagneticdissipationforce1997,KisielSuppressionelectronicfriction2011,CockinsExcitedStateSpectroscopyIndividual2012}. 
The capability of nearly symmetric NWs to transduce lateral forces in two directions allows for imaging the full vectorial character of force fields. In particular, this enables the distinction of non-conservative forces, such as optical or frictional forces, from conservative ones, such as those arising from an electrical potential. Functionalized
with a magnetic tip, a NW cantilever forms an excellent probe for the
subtle magnetic field patterns produced by nanometer-scale
magnetization textures such as domain walls, vortices, and skyrmions;
superconducting vortices; mesoscopic transport in two-dimensional
systems; and small ensembles of nuclear spins. Although NW force
sensors are just starting to be used for many of these applications,
they have already been applied as the active sensors in some of the
most sensitive and highest resolution magnetic resonance imaging
experiments realized to
date~\cite{NicholNanomechanicaldetectionnuclear2012,NicholNanoscaleFourierTransformMagnetic2013,RoseHighResolutionNanoscaleSolidState2018}.

This review is structured as follows: \cref{Sec:mechanics} gives an introduction to the mechanical dynamics of NW cantilevers. We first describe in \cref{Sec:mechanics_singlemode} flexural motion in a single oscillation direction and with displacement amplitudes small enough such that motion remains linear. Next, in \cref{Sec:mechanics_twomodes} we consider flexural motion in two different oscillation directions and the effects of vectorial forces and spatial force derivatives. This provides the conceptual background for the detection of 2D forces and force derivatives. We then see that shear force derivatives can lead to coupling between oscillations in two directions. As shown in \cref{Sec:mechanics_coherent}, such coupling can be used to implement coherent two-mode dynamics, reminiscent of the dynamics of a quantum two-level system. Finally, we treat non-linear regimes of motion in \cref{Sec:mechanics_nonlinear}. In \cref{Sec:forcesesensing_sensitivity} we discuss the high force sensitivity of NW cantilevers. In particular, we cover how the geometry of a NW impacts its force sensitivity. In \cref{Sec:forcesesensing_displacement} we discuss optical methods of detecting NW displacement in two directions. In \cref{Sec:forces} we give an overview of experimental progress thus far and of the capabilities of NW transducers for specific forces. We conclude the review with an outlook for NW force transducers in \cref{Sec:outlook}.

\section{Mechanics of nanowire cantilevers}
\label{Sec:mechanics}
\begin{table*}[t]
	\centering
	\small
	\begin{tabular}
		{l l l l l l l l}
		\hline
		\T Material & Cross-section & $d$ (nm) & $L$ (\SIUnitSymbolMicro m)  & $\omega/2\pi$ (kHz) & $k$ (N/m) & $Q$ & Ref.\B\\
		\hline
		\hline
		\T GaAs/AlGaAs & Hexagonal & 350 & 25 & 417 & 1$\cdot\text{10}^{\text{-2}}$ & 50,000 (\SI{4}{\kelvin}) & \cite{RossiVectorialscanningforce2017, BraakmanCoherentTwoModeDynamics2018} \\
		GaAs & Hexagonal & 234 & 16.8 & 598 & 8.3$\cdot\text{10}^{\text{-3}}$ & 46,553 (\SI{4}{\kelvin}) & \cite{RossiMagneticforcesensing2018a} \\
		GaAs & Hexagonal & 100 & $<$25 & 1,197 & 8.3$\cdot\text{10}^{\text{-3}}$ & 4,900 (RT) & \cite{CadedduTimeResolvedNonlinearCoupling2016} \\
		GaAs/AlGaAs & Hexagonal & 390 & 20 & 795 & $9\cdot\text{10}^{\text{-2}}$ & 6,700 (4K) & \cite{MontinaroQuantumDotOptoMechanics2014}\\
		GaAs & Hexagonal & 130 & 14.5 & 465 & - & 2,000-3,000 (RT) & \cite{FosterTuningNonlinearMechanical2016}\\
		InAs & Hexagonal & 60-80 & 4-5.5 & 2,023.9 & $3.6\cdot\text{10}^{\text{-3}}$ & 1,752 (RT) & \cite{PairisShotnoiselimited2018} \\
		SiC & Circular & 150 & 52 & 113 & 4$\cdot\text{10}^{\text{-4}}$ & 2,890 (RT) & \cite{GloppeBidimensionalnanooptomechanicstopological2014} \\
		SiC & Circular & 200 & 50 & 78 & 1.5$\cdot\text{10}^{\text{-4}}$ & 1,000 (RT) & \cite{deLepinayuniversalultrasensitivevectorial2017a} \\
		SiC & Circular & 120 & 165 & 6.7 & 3$\cdot\text{10}^{\text{-6}}$ & 3,000 (RT) & \cite{LepinayEigenmodeorthogonalitybreaking2018} \\
		SiC & Circular & 284 & 128 & 33 & - & 36,000 (RT) & \cite{PerisanuHighfactormechanical2007}\\
		SiC & Circular & 206 & 93 & 43 & - & 159,000 (RT) & \cite{PerisanuHighfactormechanical2007}\\
		SiC & Circular & 50 & 7 & 1,519 & - & 2,500 (RT) & \cite{PerisanulinearDuffingregimes2010}\\
		SiC & Circular & 300 & 6 & 6,140 & 1.5 & 33 (RT, Air)& \cite{PigeauObservationphononicMollow2015}\\
		Si & Circular & 44 & 14.4 & 210.5 & 2.8$\cdot\text{10}^{\text{-5}}$ & 9,250 (RT) & \cite{NicholDisplacementdetectionsilicon2008} \\
		Si & Circular & 46 & 12.9 & 273 & 6.6$\cdot\text{10}^{\text{-5}}$ & 7,250 (RT) & \cite{NicholDisplacementdetectionsilicon2008} \\
		Si & - & 35 & 15 & 1,060 & 6.5$\cdot\text{10}^{\text{-4}}$ & 25,000 (\SI{8}{\kelvin}) & \cite{NicholNanomechanicaldetectionnuclear2012} \\
		Si & - & 50 & 15 & 333 & 1.5$\cdot\text{10}^{\text{-4}}$ & 18,000 (\SI{6}{\kelvin}) & \cite{NicholNanoscaleFourierTransformMagnetic2013} \\
		Si & Circular & 50 & 15 & 197.5 & 2.0$\cdot\text{10}^{\text{-5}}$ & 3,000 - 3,500 (RT) & \cite{NicholControllingnonlinearitysilicon2009}\\
		Si & Elongated circular & 60, 80 & 20 & 342 & $1\cdot\text{10}^{\text{-4}}$ & 8,150 (4K) & \cite{RoseHighResolutionNanoscaleSolidState2018}\\
		Si & Hexagonal & 100-300 & 5-10 & 2,000-6,000 & - & 2,000 (RT) & \cite{Gil-SantosNanomechanicalmasssensing2010}\\
		Si & Hexagonal & 165 & 12.7 & 1,772.4 & - & 3,000 (RT) & \cite{Gil-SantosOpticalbackactionsilicon2013}\\
		Si & Hexagonal & 90 & 9.3 & 2,504.3 & - & 3,000 (RT) & \cite{Gil-SantosOpticalbackactionsilicon2013}\\
		Si & Hexagonal & 100-200 & 6-8 & 3,500 - 4,000 & $2.4 - 5\cdot\text{10}^{\text{-2}}$ & 3,000-3,500 (RT) & \cite{RamosOptomechanicsSiliconNanowires2012}\\
		Si & Hexagonal & 150 (clamp), 60 (tip) & 11.3 & 2,480 & - & - & \cite{RamosSiliconnanowireswhere2013}\\
		Si & Hexagonal & 39-400 & 2-20 & 1,000 - 12,000 & - & 3,000-25,000 (RT) & \cite{BelovMechanicalresonanceclamped2008}\\
		Si (metallized) & Hexagonal & 142 & 2.25 & 200,000 & 110.3 & 2,000 (\SI{25}{\kelvin}) & \cite{FengVeryHighFrequency2007}\\
		Si (metallized )& Hexagonal & 118 & 2.1 & 188,000 & 62.9 & 2,500 (\SI{25}{\kelvin}) & \cite{FengVeryHighFrequency2007}\\
		Si & Hexagonal & 81 & 1.69 & 215,000 & 31.4 & 5,750 (\SI{25}{\kelvin}) & \cite{FengVeryHighFrequency2007}\\
		Si & Hexagonal & 74 & 2.77 & 80,000 & 6.0 & 13,100 (\SI{25}{\kelvin}) & \cite{FengVeryHighFrequency2007}\\
		CNT & Circular & 50 & 18 & 270 & $\sim\text{10}^{\text{-4}}$ & 250 (RT) & \cite{SiriaElectronbeamdetection2017} \\
		CNT & Circular & 1 - 3 & 5 & 38,178.5 & $4.5\cdot\text{10}^{\text{-8}}$ & 2,245 (RT) & \cite{TavernarakisOptomechanicshybridcarbon2018} \\
		CNT & Circular & 4 & 1.2 & 5,955 & $2.1\cdot\text{10}^{\text{-5}}$ & - & \cite{TsioutsiosRealTimeMeasurementNanotube2017} \\
		CNT & Circular & - & - & 363.5 & $4.8\cdot\text{10}^{\text{-6}}$ & 571 (RT) & \cite{TsioutsiosRealTimeMeasurementNanotube2017}\B\\
%
%
%
		\hline
	\end{tabular}
	\caption{Experimentally determined parameters of singly-clamped NW cantilevers. Here the diameter $d$ is the average cross-sectional width and $\omega/2\pi$ is the average frequency of the fundamental flexural mode doublet. The quality factor $Q$ is the average value of the fundamental flexural mode doublet, taken for a freestanding NW, far from any sample surface and measured at low ambient pressures. RT stands for room temperature.\vspace{1em}}
	\label{table:NWs}
\end{table*}

In this section, we give an overview of the mechanical behavior of NW cantilevers. Both the geometry and the material composition play an important role in determining the mechanical response of a NW cantilever to externally applied forces and force gradients, through its displacement profile, resonance frequency, and energy dissipation. NWs differ from conventional cantilevers in possessing a nearly symmetric cross-section. As we will see, this property implies that it is possible to use NW cantilevers as bidimensional force sensors. It is therefore of key importance to carefully choose the right type of NW for a specific force sensing application. Fortunately, a wide variety of NWs has become available in the last decades, produced either through bottom-up methods or by a top-down approach. This offers researchers a choice of many materials, heterostructures combining different materials, and various types of functionalization such as integrated quantum dots or magnetic tips. Furthermore, fine control over the shape of NWs has been demonstrated, yielding NWs with a tunable diameter, length, and with a variety of different cross-sections. In \cref{table:NWs} we list experimentally determined mechanical properties of various NW cantilevers found in literature.

In the following, we model the NW cantilever as a singly-clamped beam of length $L$ and with a uniform cross-section along its length. \cref{fig:mechanics}a schematically indicates the relevant parameters descibing a singly-clamped beam undergoing flexural motion in a plane. For the purposes of force sensing, we are primarily interested in flexural modes and we do not consider longitudinal and torsional vibrations, since these modes are comparatively stiff and consequently have much higher oscillation frequencies.\\

\subsection{Linear motion of a single mode}
\label{Sec:mechanics_singlemode}
The flexural eigenfrequencies and eigenmodes of a singly-clamped beam can be derived analytically using Euler-Bernoulli theory, which gives a valid approximation when $L$ is much larger than the cross-sectional dimension and when rotational bending and shear deformation can be neglected\cite{ClelandFoundationsNanomechanicsSolidState2003}. For a beam with completely symmetric cross-section, the bending moment of inertia is independent of the oscillation direction. As a consequence, flexural motion takes place in an arbitrary direction. For very asymmetric beams, such as conventional cantilever force transducers, which are flat, thin, and long, the predominant flexural motion occurs in one direction. In both cases, we can approximate the flexural motion by considering motion in one dimension. In \cref{Sec:mechanics_twomodes}, we consider NWs with slightly asymmetric cross-sections, in which case we will generalize the formalism to take into account a second dimension of flexural motion.\\

As a starting point, we consider an idealized situation in which there is no dissipation and we do not take any driving forces into account. In this case, one can approximate the effect of a transverse force to be a torque perpendicular to both force and beam axis. An evaluation of the internal forces and torques along a flexing beam yields the following equation of motion:
\begin{equation}
E_YI\frac{\partial^4U(z,t)}{\partial z^4} + \rho A \frac{\partial^2 U(z,t)}{\partial t^2} = 0,
\label{EQ:wavefunction}
\end{equation}
with $z$ referring to the position along the beam, $E_Y$ the Young's modulus of elasticity, $I$ the second moment of area, which depends on the direction of flexural motion with respect to the cross-section, $\rho$ the mass density, and $A$ the cross-sectional area. This equation can be solved for the beam displacement $U(z,t)$, which can be separated into position- and time-dependent parts such that $U(z,t) = \sum\limits_{n=1}^{\infty}r_n u_n(z)e^{i\omega_{n} t}$. Here we sum over the $n$ particular solutions, or modes, with $u_n(z)$, $r_n$, and $\omega_n$ the one-dimensional shape, amplitude, and eigenfrequency associated with the $n^{th}$ mode, respectively. In order for $r_n$ to correspond to the amplitude of cantilever displacement, we use the normalization $|u_n(L)|=1$. Note that this normalization entails that the $n$ modes are orthogonal but not orthonormal.\\

%
Using the boundary conditions of a cantilever, i.e. that one end of the beam is fixed ($u_n(0)=0$) and not bending ($u_n'(0)=0$)) and that the other end is free to move (zero torque: $u_n''(L)=0$ and zero transverse force: $u_n'''(L)=0$),  for small displacements Euler-Bernoulli theory yields the mode shapes:
\begin{align}
u_n(z) = &A\Bigg[\bigg((\cos{(\frac{\beta_n}{L}z)} - \cosh{(\frac{\beta_n}{L}z)}\bigg) \nonumber \\ &+\frac{\cos{\beta_n}+\cosh{\beta_n}}{\sin{\beta_n}+\sinh{\beta_n}}\bigg((\sinh{(\frac{\beta_n}{L}z)} - \sin{(\frac{\beta_n}{L}z)}\bigg)\Bigg],
\label{EQ:modeshape}
\end{align}
with $A$ a normalization constant and $\beta_n$ the dimensionless wavenumber of mode $n$. \cref{fig:mechanics}b shows the first three of these flexural mode shapes. The modes $u_n(z)$ can now be used in \cref{EQ:wavefunction}, resulting in the eigenfrequencies:
\begin{equation}
\omega_n = \frac{\beta_n^2}{L^2}\sqrt{\frac{E_YI}{\rho A}}
\label{EQ:eigenfrequencies}
\end{equation}
\cref{table:betan} lists values of $\beta_n$ and normalized values of the eigenfrequencies.
\begin{table}[b]
	\begin{tabular}
		{l l l}
		$n$ & $\beta_n$ & $\omega_n/\omega_0$\\
		\hline
0 & 1.875 & 1.000\\
1 & 4.694 & 6.267\\
2 & 7.855 & 17.547\\
3 & 10.996 & 34.386\\
$n\geq3$ & $(n+1/2)\pi$ & [$(n+1/2)\pi/\beta_0]^2$\\
		
		\hline
	\end{tabular}
	\caption{Wave numbers $\beta_n$ and normalized eigenfrequencies $\omega_n$ of a singly-clamped beam.\vspace{1em}}
	\label{table:betan}
\end{table}

The eigenmodes and eigenfrequencies of \cref{EQ:modeshape} and \cref{EQ:eigenfrequencies} correspond to the ideal case of a cantilever without dissipation and without specifying any external transverse forces. To describe realistic time-dependent behavior of the beam, we must also include dissipation and external driving forces. Dissipation can be characterized by a quality factor, which is defined as the ratio between the total stored energy of a resonator and the energy loss per cycle: $Q = 2\pi E/\Delta E$. In terms of the displacement, $Q$ is a dimensionless constant describing exponential decay:
$U(z,t) = \sum\limits_{n=1}^{\infty}u_n(z)e^{i\omega_{0n} t}e^{-\omega_{0n} t/(2Q)} = \sum\limits_{n=1}^{\infty}U_n(z)e^{i(1 + i/(2Q)\omega_{0n}t}$. Hence, we see that dissipation can be incorporated by defining a new eigenfrequency $\omega_{0n}'= (1 + i/2Q)\omega_{0n}$.\\



Mechanical force sensors sensitively transduce forces and force gradients into a change of displacement or resonance frequency. To explain this in more detail, we now introduce transverse driving forces of the form $F_n(z,t) = F_{0n}(z)e^{i\omega_n t}$. For times long compared to transient time scales ($t>>Q_n/\omega_n$), the displacement of the beam follows the driving force and has the form $U(z,t) = \sum\limits_{n=1}^{\infty}U_n(z)e^{i\omega_n t}$. In the following, we only consider the motion of the free end ($z=L$) of the cantilever. By adding $F_n(L,t)$ to the right-hand-side of \cref{EQ:wavefunction}, we can evaluate the complex displacement response of the free end of the cantilever as a function of driving frequency $\omega_n$ as:
\begin{equation}
r_n(\omega_n) = F_n(\omega_n)\chi_n(\omega_n) \equiv \frac{F_n(\omega_n)}{m_{n,eff}}\frac{1}{\omega_{0n}^2 - \omega_n^2 + i\omega_{0n}^2/Q_n}
\label{EQ:response}
\end{equation}
Here $F_n(\omega_n)$ is the Fourier transform of $F_n(L,t)$ and $\chi_n(\omega_n)$ the mechanical susceptibility of the NW cantilever. Note that the displacement response given by the mode shapes requires the use of a mode-dependent effective mass $m_{n,eff}$ rather than the cantilever mass in describing the dynamical behavior of the cantilever. This can be udnerstood by considering that elements of the beam located closer to its clamped end react to a transverse force as if the local mass were higher than for elements closer to the free end. In fact, the effective mass is proportional to the square of the mode volume $m_{n,eff} = \frac{1}{|u_n(z_0)|^2}\int \rho(z) |u_n(z)|^2 dV$ and hence depends on the chosen normalization of $u_n(z)$~\cite{PootMechanicalsystemsquantum2012}. Furthermore, $m_{n,eff}$ depends on the position along the beam $z_0$ for which motion is evaluated. For the fundamental flexural mode using our normalization condition $|u_n(L)| = 1$, it has a minimum of $m_{0,eff} = M/4$ for $z_0 = L$, with $M$ the cantilever mass. From here on we will use the notation $m\equiv m_{0,eff}$. Interestingly, \cref{EQ:response} differs only slightly from the response of a simple harmonic oscillator, for which $i\omega_{0n}^2/Q_n$ is replaced by $i\omega_n\omega_{0n}/Q_n$. The difference is negligible for large values of $Q_n$ and disappears when driving on resonance. 

In the following, we will therefore approximate the dynamics of a mode of a singly-clamped NW cantilever with that of a driven damped harmonic oscillator, considering only the displacement of the free end of the NW and assuming near-resonant driving.
%
%
%
For simplicity, from now, we only consider the fundamental flexural mode (typically the only one used in force sensing) and consequently we drop the subscript $n$. For small NW oscillation amplitudes, we can approximate the force $F$ acting on the NW around its equilibrium displacement $r=0$ as $F \approx F(0)+rF_r$, with $F_r \equiv \partial F / \partial r|_0$. Writing the displacement $r(t)$ of the tip of the nanowire in an equation of motion of a damped driven harmonic oscillator, we obtain:
\begin{equation}
m\ddot r + \Gamma\dot r + (k - F_r)r = F(0) + F_{Th},
\label{EQ:EOM1}
\end{equation}
where $k$ is the mode spring constant and dots indicate differentiation with respect to time. We see that spatial derivatives $F_r$ of forces acting on the NW modify the effective spring constant of the NW. Stationary forces acting on the NW provide a static offset of the cantilever displacement, whereas time-dependent forces can drive the motion of the NW and change its displacement amplitude, through the spectral transfer function of the resonance shown in \cref{EQ:response}. The aim of most force sensing experiments is the measurement of the forces $F(0)$ and force derivatives $F_{r}$.\\ 

The remaining terms in \cref{EQ:EOM1}, $\Gamma\dot r$ and $F_{Th}$, correspond to dissipation and thermal force noise, respectively. Forces acting on the resonator typically do not only arise from controlled interactions, but also include components originating from the cantilever being in contact with a thermal bath. The noise in the resonator motion that results from this interaction is observable in the most sensitive force sensing measurements, and is what typically sets the limit to their sensitivity. We include further terms in the equation of motion that account for energy dissipation and noise arising from the resonator being in contact with a largely uncontrolled and fluctuating environment of temperature $T$. This environment consists of a large number of degrees of freedom, such as those describing the motion of the molecules in the (typically highly diluted) fluid through which the nanowire moves and the phonon bath in the substrate to which the NW is attached through its clamping point. We use a statistical approach to include these effects in the equation of motion. The interactions with the environment can be decomposed into a rapidly fluctuating thermal force noise term $F_{Th}$ and a slowly varying dissipation term proportional to the velocity of the resonator, $\Gamma\dot{r}$ (see Reif~\cite{ReifFundamentalsstatisticalthermal1965}), with $\Gamma = m\omega/Q$. While in thermal equilibrium, these two terms are related through the fluctuation-dissipation theorem~\cite{Kubofluctuationdissipationtheorem1966}, resulting in:
\begin{equation}
\Gamma = \frac{1}{2k_BT}\int\limits_{-\infty}^{\infty}\langle F_{Th}(t)F_{Th}(t+s)\rangle ds,
\label{EQ:fluctdiss}
\end{equation}
where $k_B$ is the Boltzmann constant and the brackets $\langle...\rangle$ indicate that an ensemble average is taken, here used in the correlation function $\langle F_{Th}(t)F_{Th}(t+s)\rangle$, with $t$ and $t+s$ different points in time. To further describe the fluctuating force term $F_{Th}(t)$, we use the power spectral density (PSD). We define the force PSD $S_F(\omega)$ as the Fourier transform of the force correlation function:
\begin{equation}
S_F(\omega) = 2\int\limits_{-\infty}^{\infty}\langle F(t)F(t+\tau)\rangle e^{i\omega\tau}d\tau, \text{ for } 0 < \omega < \infty
\label{EQ:PSDSF}
\end{equation}
Here we defined $S_F(\omega)$ to be single-sided: since $\langle F_{Th}(t)F_{Th}(t+s)\rangle$ is real, its Fourier transform is even. This allows us to only take into account positive values of $\omega$ in \cref{EQ:PSDSF}, provided we multiply the integral by 2. Furthermore, note that from the inverse Fourier transform of \cref{EQ:PSDSF} we can find specifically that $\langle F(t)^2\rangle =\frac{1}{2\pi}\int\limits_{0}^{\infty} S_F(\omega)d\omega$.
Correlations in a thermal bath (other than those taken into account for damping) usually exist only for extremely short times (typically $< 10^{-13}$\,s). Therefore, for thermal noise we can assume $\langle F_{Th}(t)F_{Th}(t+s)\rangle = 2\Gamma k_B T\delta(s)$, resulting in the white single-sided force PSD $S_{F,Th}=4k_BT\Gamma$.

\begin{figure}[t]
	\includegraphics[width=0.45\textwidth]{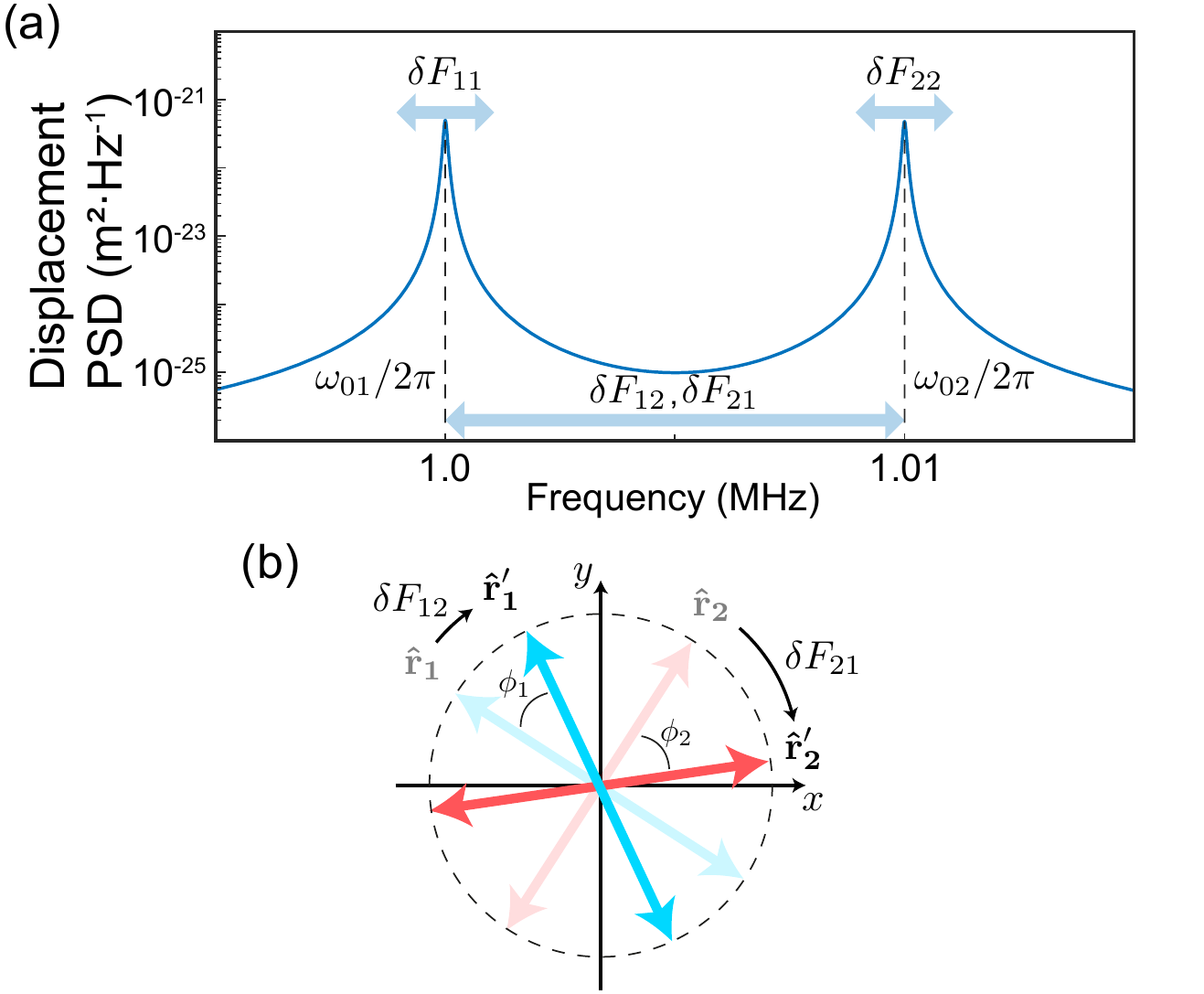}
	\caption{(a) Plot of thermal displacement PSD of a nearly-degenerate mode doublet. Arrows indicate the effect of the force derivatives $F_{ij}$ on the resonant frequencies. (b) Schematic picture of rotation of mode directions under the influence of off-diagonal force derivatives $F_{12}$ and $F_{21}$. Modes 1, 2 oscillate in the directions of the unit vectors $\hat{\mathbf{r}}\mathbf{_{1,2}}$. The force derivatives $\delta F_{12}$ and $\delta F_{21}$ rotate the mode oscillation directions to point along the new unit vectors $\hat{\mathbf{r}}'\mathbf{_{1,2}}$.\vspace{1em}}
	\label{fig:twomodes}
\end{figure}
The single-sided displacement PSD $S_r(\omega)$ can be similarly expressed by combining the amplitude response function of \cref{EQ:response} with \cref{EQ:PSDSF} as:
\begin{align}
S_r(\omega) &= 2\int\limits_{-\infty}^{\infty}\langle r(t)r(t+\tau)\rangle e^{i\omega\tau}d\tau\nonumber, \text{ for } 0 < \omega < \infty\\
 &=S_F(\omega)|\chi|^2(\omega) = \frac{S_F(\omega)}{m^2}\frac{1}{(\omega_0^2 - \omega^2)^2 + \omega_0^4/Q^2}
\label{EQ:PSDamp}
\end{align}
We see that the mechanical resonator acts as a filter on the driving force, with $\chi(\omega)$ the filter transfer function. An important example of this filtering effect is that spectrally white force noise results in displacement noise with a nearly Lorentzian spectrum with a maximum of $S_{r,Th} = 4k_BTQ/m\omega_0^3$. As discussed in \cref{Sec:forcesesensing_sensitivity}, this thermally induced displacement noise is what ultimately limits the sensitivity of force detection. Finally, we note that thermally induced phase noise (see Refs.\cite{RobinsPhasenoisesignal1982, ClelandFoundationsNanomechanicsSolidState2003} for a thorough treatment) similarly limits the frequency stability of a mechanical resonator\cite{SansaFrequencyfluctuationssilicon2016}.



\subsection{Linear motion of two orthogonal modes}
\label{Sec:mechanics_twomodes}
For any cross-section that is symmetric around its center point, the moments of inertia $I_x$ and $I_y$ are equal, resulting in degenerate eigenmodes. The flexural oscillation direction in the plane is arbitrary in this case. Then the formalism of a single mode direction as discussed in the previous subsection suffices to describe the motion. However, small asymmetries in the cross-section or in the clamping conditions~\cite{NicholDisplacementdetectionsilicon2008} lift this degeneracy and split each mode into a doublet of modes oscillating along orthogonal directions $\hat{\mathbf{r}}\mathbf{_{1}}$ and $\hat{\mathbf{r}}\mathbf{_{2}}$ (see \cref{fig:twomodes}). For the fundamental flexural mode of NW cantilevers, this results in a doublet with resonance frequencies that are typically split by a small fraction of the resonance frequencies, but many times the resonance linewidths. Several groups working with NW force sensors have observed orthogonal modes with frequency splittings of the order of $10^3-10^4$\,Hz. Moreover, similar orthogonal flexural modes can arise in doubly-clamped beams and have been observed, for example, in carbon nanotubes and SiN etched beams~\cite{MoserUltrasensitiveforcedetection2013,FaustNonadiabaticDynamicsTwo2012,FaustCoherentcontrolclassical2013}. The use of such mode doublets as a way to realize bidimensional force sensing forms a major motivation for using NW cantilevers as force transducers and we will treat it in detail in this review. The mechanical dynamics of these mode doublets are very similar to that of two coupled modes and we can therefore apply the formalism of the previous subsection generalized to two modes.\\

Taking into account both orthogonal modes, the equation of motion can now be written in vectorial form as:
\begin{equation}
m\ddot{\textbf{r}} + \mathbf{\Gamma}\cdot\dot{\textbf{r}} + \textbf{K}\cdot\textbf{r} = \textbf{F}(0) + \textbf{F}_{Th}
\label{EQ:EOM2}
\end{equation}
Here the displacement and forces are vectors defined in the basis of the two mode directions $\hat{\mathbf{r}}\mathbf{_1}$ and $\hat{\mathbf{r}}\mathbf{_2}$, i.e. $\textbf{r} \equiv \left(\begin{smallmatrix}r_1\\r_2\end{smallmatrix}\right)$, $\textbf{F}(0) \equiv \left(\begin{smallmatrix}F_1(0)\\F_2(0)\end{smallmatrix}\right)$, and $\textbf{F}_{Th} \equiv \left(\begin{smallmatrix}F_{Th,1}\\F_{Th,2}\end{smallmatrix}\right)$. The effective mass $m$ is taken to be equal for the two modes (see for experimental verification of this the Supplementary Information of Ref.\cite{GloppeBidimensionalnanooptomechanicstopological2014}), and $r_i$ are the mode displacements. $F_{Th,i}$ and $F_i$ represent thermal force noise and other external forces acting on each of the two modes, respectively. 
We use the small-displacement approximation around $r_i=0$, $F_i \approx F_i(0)+r_j\frac{F_i}{\partial r_j}|_{0}$. Furthermore, we define the dissipation and effective spring constant matrices $\mathbf{\Gamma} \equiv \left(\begin{smallmatrix} \Gamma_1 & 0 \\ 0 & \Gamma_2 \end{smallmatrix} \right)$ and $\textbf{K} \equiv \left(\begin{smallmatrix} k_1 - F_{11} & - F_{21} \\ - F_{12} & k_2 - F_{22} \end{smallmatrix} \right)$. We write the four spatial force derivatives in shorthand notation as $F_{ij} \equiv \frac{\partial F_i}{\partial r_j}|_{0}$, with $i,j\in \{1,2\}$. 

We see that the force derivatives $F_{ii}$ cause a change in the spring constant of each mode, similar to the single-mode case. The off-diagonal elements of $\textbf{K}$ linearly couple the two modes, resulting in two new hybridized modes. The eigenvalues of the hybridized modes obtained by diagonalizing the $\textbf{K}$ matrix can be expressed as the modified spring constants (for $\Gamma_i/2m << \sqrt{k_i/m}$):
\begin{align}
k_{1,2}' = &\frac{1}{2} \bigg [k_1 + k_2 - F_{11} - F_{22}\nonumber \\
&\pm \sqrt{\Big(k_1 - k_2 -F_{11} + F_{22}\Big)^2 + 4 F_{12}F_{21}}  \bigg ].
\label{Eq:newspringconstants}
\end{align}
The modified eigenfrequencies then follow as $\omega_{0,i}' = \sqrt{k_i'/m}$. The directions of the corresponding new eigenmodes can be written as:
\begin{align}
&\hat{\mathbf{r}}'\mathbf{_{1}} = \frac{1}{\sqrt{(k_2-F_{22}-k_1')^2 + F_{12}^2}}\begin{pmatrix}
k_2-F_{22}-k_1'\\
F_{12}
\end{pmatrix},\nonumber \\
&\hat{\mathbf{r}}'\mathbf{_{2}} = \frac{1}{\sqrt{(k_1-F_{11}-k_2')^2 + F_{21}^2}}\begin{pmatrix}
F_{21}\\
k_1-F_{11}-k_2'
\end{pmatrix}
\label{Eq:neweigenmodes}
\end{align}
We can further simplify \cref{Eq:newspringconstants} by making the assumption that the force derivatives are much smaller than the bare NW spring constants $k_i$, which is usually satisfied. This yields $k_i'\approx k_i - F_{ii}$ and the following expression for the diagonal elements of the force derivative matrix:
\begin{equation}
F_{ii}|_{r_i = 0}\approx -2k_i\Delta\omega_{0,i}/\omega_{0,i}.
\label{EQ:Fii}
\end{equation}
Here $\Delta\omega_{0,i}$ is the frequency shift of mode $i$ induced by the force derivatives. Using the same approximation, the off-diagonal elements of the force derivative matrix (shear force derivatives) can be deduced from the angle $\phi_i$ (see \cref{fig:twomodes}b)between the mode direction $\hat{\mathbf{r}}'\mathbf{_{i}}$ in the presence of the force derivatives and the bare mode direction $\hat{\mathbf{r}}\mathbf{_{i}}$ to be:
\begin{equation}
F_{ij}|_{r_i = 0}\approx |k_i - k_j|\tan({\phi_i}) \textrm{   for  }i\neq j.
\label{EQ:Fij}
\end{equation}
Note that in the situation where the NW cantilever is very soft and strong force derivatives are probed, these approximations are not valid and expressions for $F_{ii}$ and $F_{ij}$ remain slightly more complicated. In this case, the effects of the external force field dominate over the intrisic NW mechanical properties, and the new eigenmodes tend to align with the eigenvectors of the force derivative matrix\cite{deLepinayuniversalultrasensitivevectorial2017a}.

\cref{EQ:Fii} and \cref{EQ:Fij} show that all in-plane force derivatives can be determined from measurements of the frequencies and oscillation angle of the two modes. The effect of $F_{ij}$ on the two resonance frequencies is illustrated in \cref{fig:twomodes}a. Finally, we note that the hybridized modes remain orthogonal as long as only conservative forces are included, in which case $F_{12} = F_{21}$. For forces with a finite curl, such as those arising from optical or frictional forces, $F_{12} \neq F_{21}$ and orthogonality is not maintained. In this situation, the standard fluctuation-dissipation theorem is no longer a valid method of describing the interactions of the bimodal NW with a thermal environment and needs to be modified to include excess noise arising from the non-zero curl of $\textbf{F}$~\cite{LepinayEigenmodeorthogonalitybreaking2018}.
An important further consequence of non-orthogonality is that it becomes impossible to distinguish a rotation of a mode from a change in mode temperature without a full vectorial read-out as described in \cref{Sec:forcesesensing_displacement}.

\begin{figure*}[t]
	\includegraphics[width=0.95\textwidth]{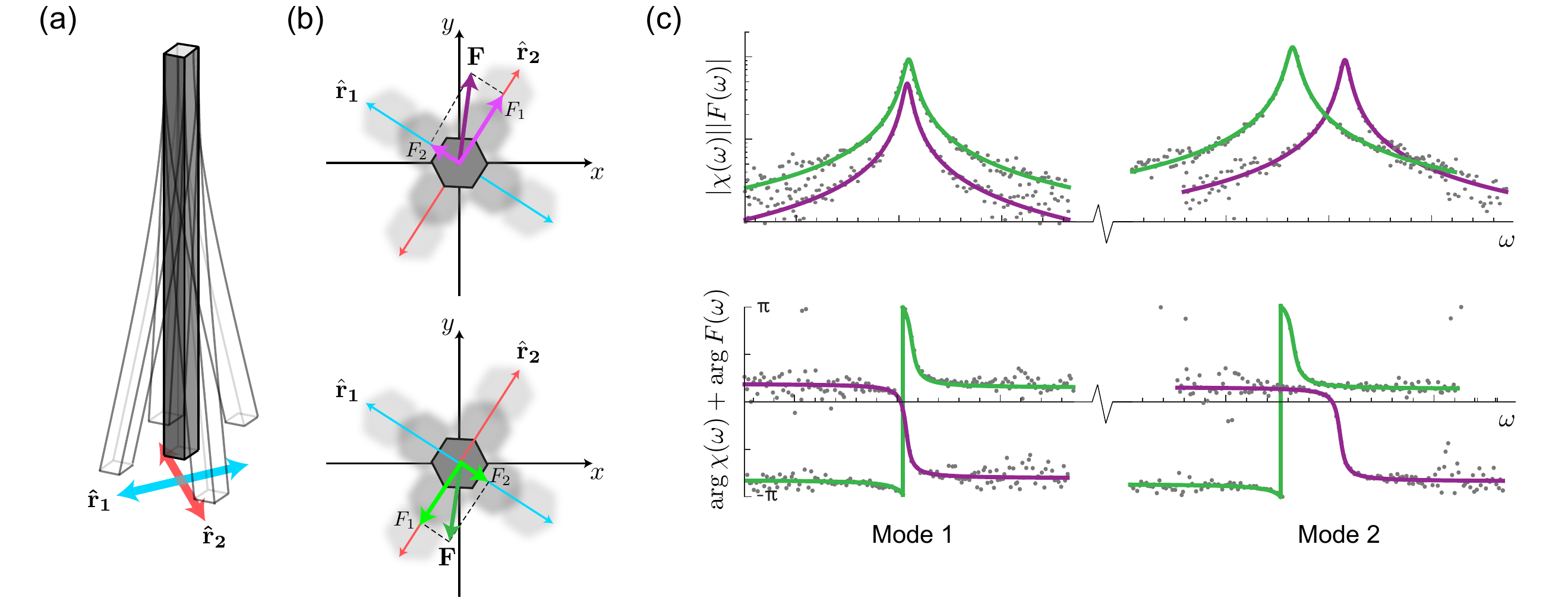}
	\caption{(a) Illustration of two orthogonal flexural modes in a NW cantilever. (b) Illustration of projection of vectorial forces onto mode directions, in the absence of spatial force derivatives. (c) Displacement amplitude and phase response of two modes as a function of driving frequency. The green and purple curves correspond to the response to the two forces indicated in (b) with the same colors. The displacement of each mode allows to extract the magnitude of the force component along that mode direction. The phase response of this mode with respect to the driving then allows to determine the sign of the force component along each mode direction. Adapted from Ref.\cite{RossiVectorialscanningforce2017}\vspace{1em}}
	\label{fig:forces}
\end{figure*}
The forces $\mathbf{F}$ and $\mathbf{F}_{Th}$ are two-dimensional generalizations of the forces acting on a single mode. As for a single mode, stationary forces $\mathbf{F_1}$ and $\mathbf{F_2}$ yield a static offset in the NW displacement, now for each mode direction. Time-dependent forces drive the modes, leading to mode displacements given by a generalization of \cref{EQ:response}:
\begin{equation}
r_i'(\omega) = \textbf{F}(\omega)\cdot\hat{\mathbf{r}}'\mathbf{_{i}}\chi_i'(\omega) \textrm{  with }i,j\in \{1,2\},
\label{EQ:vectorialresponse}
\end{equation}
with $\hat{\mathbf{r}}'\mathbf{_{i}}$ the unit vector along the oscillation direction and $\chi_i'(\omega)$ the susceptibility of mode $i$, respectively. As described before, the force derivatives acting on the NW lead to the modified spring constants $k_i'$ and mode directions $\hat{\mathbf{r}}'\mathbf{_{i}}$, resulting in the new susceptibilities $\chi_i'(\omega)$. \cref{fig:forces}b and c show how measurements of the displacement amplitudes and phases of both modes as a function of driving frequency can be used to reconstruct the full vectorial driving force.

\subsection{Coherent two-mode dynamics}
\label{Sec:mechanics_coherent}
\begin{figure}[t]
	\includegraphics[width=0.45\textwidth]{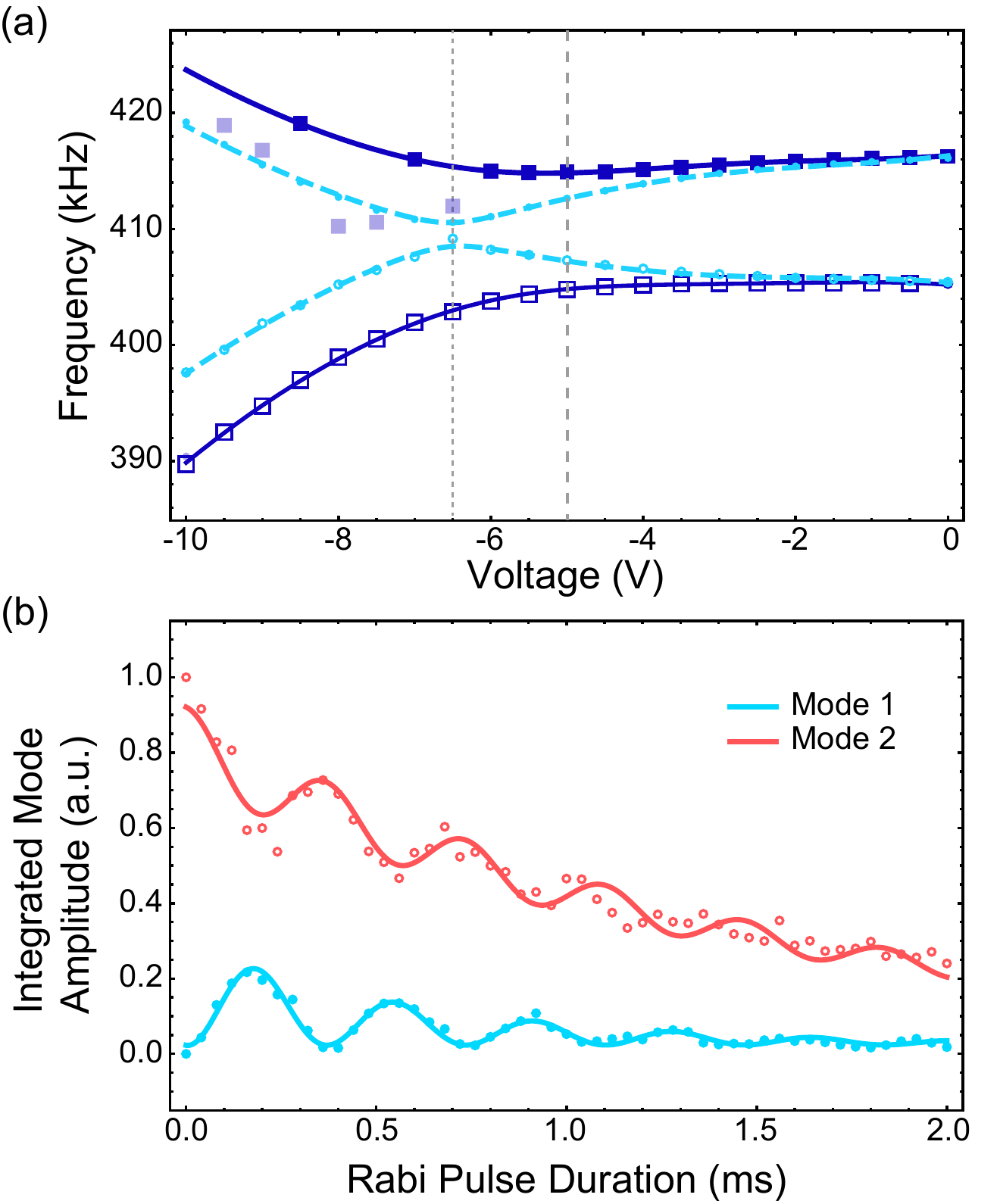}
	\caption{(a) Mode frequencies as a function of voltage applied over two nearby surface gates, for two different positions of the NW tip inside the electric field. (b) Integrated displacement amplitudes of both modes as a function of Rabi pulse duration. Points indicate measured data, solid lines are fits to the data. Note that the offset in the measured Rabi oscillations is due to thermal excitation of the mode. Different offsets result from the difference in angle of the two mode directions with the read-out vector. Adapted from Ref.\cite{BraakmanCoherentTwoModeDynamics2018}. \vspace{1em}}
	\label{fig:coherent}
\end{figure}
As mentioned in the above, the shear force derivatives $F_{ij}$, with $i\neq j$, couple the motion in the two modes. Interestingly, the resulting hybridized eigenmodes display coherent dynamics similar to that of quantum two-level systems. Coherent two-mode phenomena such as avoided mode crossings and Rabi oscillations have been observed in classical systems including modes of optical ring resonators\cite{SpreeuwClassicalrealizationstrongly1990} and recently also in nanomechanical resonators~\cite{FaustNonadiabaticDynamicsTwo2012,FaustCoherentcontrolclassical2013,OkamotoCoherentphononmanipulation2013}. Furthermore, coupling of the orthogonal fundamental flexural modes of a NW cantilever was demonstrated through the measurement of avoided crossings of the mode frequencies~\cite{BraakmanCoherentTwoModeDynamics2018}, see \cref{fig:coherent}a. In the same experiment, Rabi oscillations were demonstrated through the use of time-dependent electric field derivatives, see \cref{fig:coherent}b. The NW was embedded in a scanning probe setup, which made it possible to tune both the coupling strength and the Rabi frequency by positioning the NW inside the electric field provided by the sample. Such coherent two-mode dynamics does not only illustrate the similarities of quantum and classical wave mechanics, but can also be used to enhance force sensing. In particular, coherent pulse sequences reminiscent from quantum control techniques\cite{DegenQuantumsensing2017}, such as Hahn echoes and dynamical decoupling, give the potential to increase the frequency stability of mechanical sensors, ultimately leading to higher sensitivities. Furthermore, such pulsing techniques make it possible to implement a range of noise spectroscopy methods in classical force and mass sensing\cite{DegenQuantumsensing2017}. 

\subsection{Non-linear motion}
\label{Sec:mechanics_nonlinear}
\begin{figure}[b]
	\includegraphics[width=0.45\textwidth]{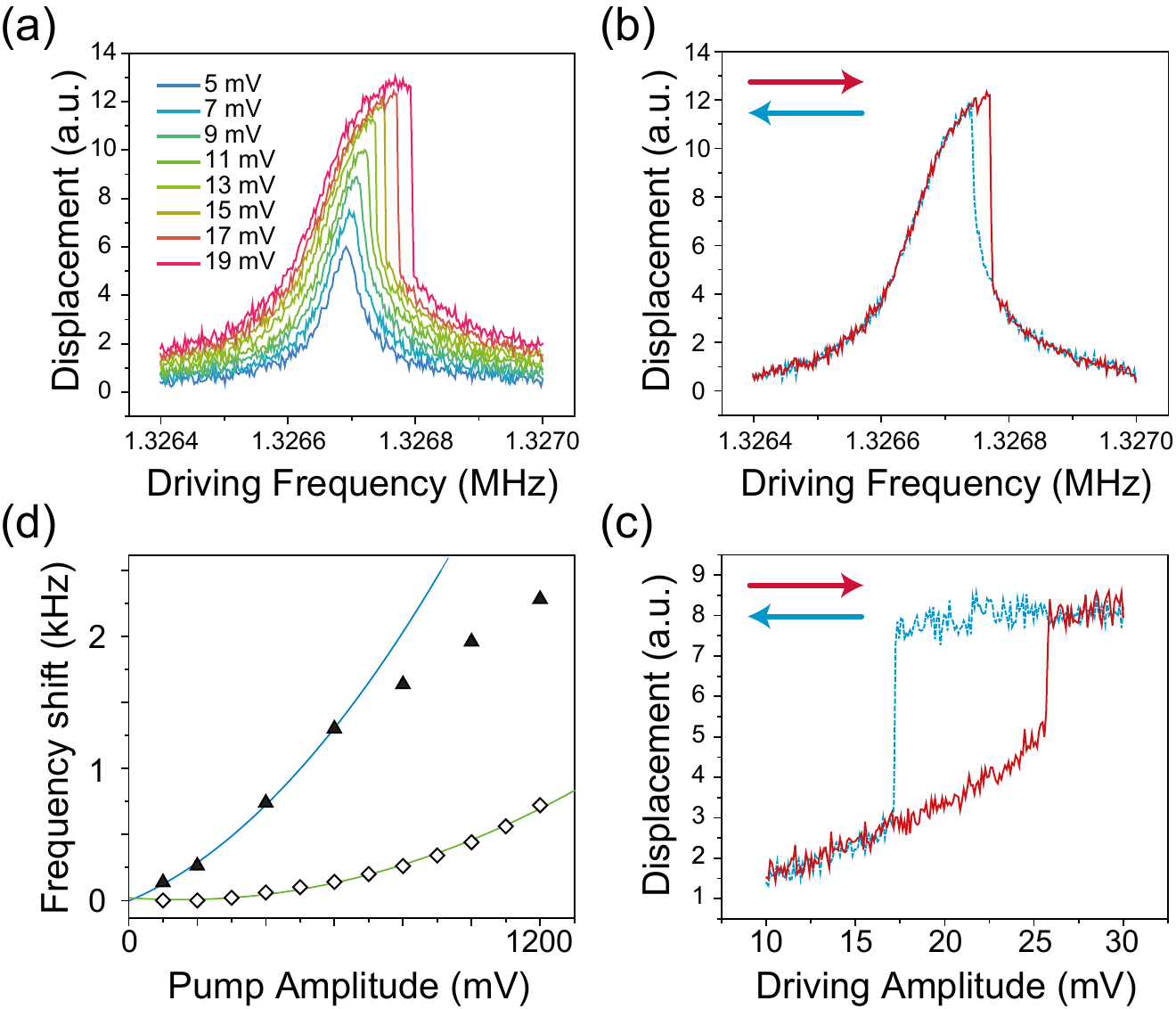}
	\caption{(a) Displacement as a function of driving frequency, for various amplitudes of piezoelectric driving. (b) Displacement as a function of driving frequency (at a driving amplitude of \SI{17}{\milli\volt}), for two sweep directions (as indicated by arrows). (c) Displacement as a function of driving amplitude (at a driving frequency of \SI{1.32677}{\mega\hertz}), for two sweep directions. (d) Frequency shift of one mode as a function of amplitude of piezoelectric driving of orthogonal mode. Adapted from Ref.\cite{BraakmanNonlinearmotionmechanical2014}.\vspace{1em}}
	\label{fig:nonlinearity}
\end{figure}
In force sensing, non-linear regimes of motion are typically avoided. However, in nanomechanical resonators suchs as NWs, non-linear motion typically arises already for modest driving amplitudes and therefore needs to be considered. Moreover, non-linearities can be put to good use, since they allow to implement amplification of forces through parametric driving\cite{ClelandFoundationsNanomechanicsSolidState2003} or through mechanical mixing\cite{BraakmanNonlinearmotionmechanical2014}. Furthermore, parametric driving allows to increase the ratio of transduced force signal to read-out noise\cite{Eichlerparametricsymmetrybreaking2018}. In the last part of this section, we treat the situation in which the displacement of the NW cannot be considered to be small anymore and non-linear terms need to be added to the equation of motion that depend on the oscillation amplitude.\\

It can be shown that for the fundamental flexural mode of a cantilever, the geometric non-linearity, associated with the lengthening of the beam as it flexes, dominates over other types of non-linearity, such as inertial non-linearity\cite{SilvaNonlinearFlexuralFlexuralTorsionalDynamics1978,SilvaNonlinearFlexuralFlexuralTorsionalDynamics1978a,SchmidFundamentalsNanomechanicalResonators2016}. The first relevant higher-order terms are a damping term, as well as an effective spring constant term, that both have a quadratic dependence on the oscillation amplitude. For a single mode, this results in the following equation of motion~\cite{CadedduNanomechanicsscanningprobe}:
\begin{equation}
m\ddot r + \Gamma\dot r + \eta r^2\dot r + kr + \alpha r^3 = F(0) + F_{Th}(0) + F_rr
\label{EQ:duffingEOM}
\end{equation}
The coefficient $\alpha$ parametrizes the strength of the cubic (Duffing) non-linearity, whereas $\eta$ is the coefficient of non-linear damping (note that here we contracted $m$ into $\alpha$ and $\eta$). For weak damping and weak anharmonicity, other second- and third-order terms can be absorbed into the coefficients $\alpha$ and $\eta$~\cite{LifshitzNonlineardynamicsnanomechanical,EichlerNonlineardampingmechanical2011}. The equation of motion \cref{EQ:duffingEOM} leads to an amplitude response as a function of driving frequency with a characteristic shark-fin shape, as shown in Figure~\ref{fig:nonlinearity}. This lineshape is a consequence of \cref{EQ:duffingEOM} having two stable solutions within a certain frequency range. This bistability leads to the switching phenomena seen for high driving amplitudes at the right flank of the response peak (\cref{fig:nonlinearity} (a)). Which of the two solutions is realized, is determined by the initial conditions, and mechanical hysteresis can be observed when adiabatically sweeping the driving frequency or driving amplitude up and down (\cref{fig:nonlinearity} (b) and (c)). When $\alpha$ is positive (negative), the Duffing non-linearity increases (decreases) the effective spring constant with increasing driving amplitude, thus stiffening (softening) the motion. The value and sign of $\alpha$ has been demonstrated to be tunable both through the shape of the NW~\cite{FosterTuningNonlinearMechanical2016}, as well as through feedback using a nearby gate electrode~\cite{NicholControllingnonlinearitysilicon2009}. The non-linear damping term has the effect of decreasing the shift of the frequency of maximum response amplitude due to the Duffing non-linearity, as well as decreasing the size of the hysteresis loop. Non-linear damping has so far not been demonstrated for NW cantilevers. Finally, note that the dependence of mode frequencies on the displacement can lead to coupling between the two orthogonal modes. Considering only the Duffing non-linearity, the non-linear term for two modes becomes $\alpha|\textbf{r}|^2\textbf{r}$; this term induces a coupling between two modes. The coupling can be observed as a quadratic shift of a mode frequency as a function of the oscillation amplitude of the other mode (see \cref{fig:nonlinearity} (d)).

Notwithstanding its potential for force amplification, non-linear motion has so far not been used in experiments on force sensing with NW cantilevers. In the following sections, we therefore only consider the operation of NW force transducers in a linear regime of motion.

\section{Force sensing with nanowire cantilevers}
\label{Sec:forcesensing}
\subsection{Force sensitivity}
\label{Sec:forcesesensing_sensitivity}
The key component in any force microscope is the force sensor. This
device consists of a mechanical transducer, used to convert force into
displacement, and an optical or electrical displacement
detector. Although early AFM transducers were simply pieces of gold or
aluminum foil~\cite{RugarAtomicForceMicroscopy2008}, specially designed and
mass-produced Si cantilevers soon became the industry standard and led
to improved resolution and force
sensitivity~\cite{AkamineImprovedatomicforce1990}. These micro-processed
devices are now cheap, readily available, and designed -- depending on
the target application -- to have a variety of features including
coatings, electrical contacts, or magnetic tips.

Conventional top-down cantilevers are well-suited for the measurement
of the large forces and force gradients present on the
atomic-scale. Nevertheless, for some applications sensitivity to small
forces is crucial. These range from mass detection, to cantilever
magnetometry, to scanning measurements of friction forces, Kelvin
probe microscopy, electric force microscopy, magnetic force microscopy
(MFM), and force-detected magnetic resonance. This push towards higher
sensitivity has generated an interest in using ever smaller mechanical
force transducers, especially those made by bottom-up
techniques. 

The trend towards decreasing the size of mechanical transducers is
based on fundamental principles. Once the detection of a transducer's
displacement is optimized, the minimum detectable force
$F_{\text{min}}$ is ultimately limited by the thermal force
fluctuations acting on the transducer, whose PSD, as discussed in
\cref{Sec:mechanics_singlemode}, is given by $S_{F,th}$.  As a result,
\begin{equation}
F_{\text{min}} = \sqrt{S_{F,th}} = \sqrt{4 k_B T \Gamma}, \label{eq1}
\end{equation}
Optimizing sensitivity therefore involves reducing the
operating temperature and the dissipation.  Note that efforts aimed
purely at increasing mechanical quality factor may not necessarly
minimize $F_{\text{min}}$.  Work on so-called `damping dilution', uses
tension applied to strings or membranes to increase resonant frequency
$\omega$ while holding $\Gamma$ constant, thereby increasing
$Q$~\cite{GonzalezBrownianmotionmass1994,CagnoliDampingdilutionfactor2000,SchmidFundamentalsNanomechanicalResonators2016,TsaturyanUltracoherentnanomechanicalresonators2017,GhadimiElasticstrainengineering2018}.
Although these methods do not improve $F_{min}$, they do reduce the thermal decoherence time,
which is important for experiments seeking to probe macroscopic
quantum superposition states~\cite{MarshallQuantumSuperpositionsMirror2003}.

The sensitivity to a number of measurements is closely related
to $F_{\text{min}}$~\cite{RugarAdventuresattonewtonforce2001}.  For example, the
minimum detectable force gradient is given by $\left ( \partial F
/ \partial x \right )_{\text{min}} = F_{\text{min}}/x_{\text{osc}}$,
where $x_{\text{osc}}$ is the root-mean-square oscillation amplitude
of the transducer.  Cantilever beams are also excellent torque
transducers and, for some applications, torque is the measurement
quantity of interest, e.g.\ in torque magnetometry.  A cantilever's
thermally limited torque sensitivity is given by $\tau_{\text{min}} =
l_e F_{\text{min}}$, where $l_e$ is the effective length of the
cantilever, which takes into account the shape of the flexural
mode~\cite{StipeMagneticDissipationFluctuations2001}.  Finally, for experiments that
measure energy dissipation, $\Gamma_{\text{min}} =
F_{\text{min}}/(\omega_0 x_{\text{osc}})$.

In practice, this means that at a given temperature, a well-designed
cantilever transducer must simultaneously have low $m \omega_0$ and
large $Q$. In the limit of long and thin cantilever beam, the
Euler-Bernoulli equations imply that $m \omega_0 \propto d^3 / l$,
where $d$ is its diameter and $l$ its length.  For sensitive
transducers, experiments show that $Q$ is limited by surface-related
losses, which lead to a linear decrease of $Q$ with increasing
surface-to-volume ratio, i.e.\ $Q \propto
d$~\cite{TaoPermanentreductiondissipation2015}. As a result, we see that $\Gamma \propto
d^2 / l$, implying that long and thin cantilevers should be the most
sensitive. In fact, a review of real transducers confirms this trend.

High mechanical resonance frequencies are also attractive for
sensitive force transducers, since they allow for the measurement of
fast dynamics and they decouple the sensor from common sources of
noise. A prominent example is the additional noise experienced by a
cantilever as its tip approaches a
surface~\cite{StipeNoncontactFrictionForce2001,KuehnDielectricFluctuationsOrigins2006}. This
so-called non-contact friction is largely due to electronic
fluctuators on the surface and typically has a $1/f$-like frequency
dependence. As a cantilever approaches a surface, $\Gamma$ usually
increases and its force resolution suffers. Such processes can be
mitigated through the use of high-frequency cantilevers. When the
resonant frequency of the mechanical oscillator is much higher than
the characteristic frequency of the external noise, the resonator can
be effectively decoupled from that noise. 

A cantilever's angular resonance frequency is given by $\omega_0
\propto d / l^2$. Therefore, if we scale each of its dimensions uniformly by a factor $\beta$, we find that $\omega_0 \propto 1 /
\beta$, while $\Gamma \propto \beta$.  Therefore, in order to
simultaneously maximize $\omega_0$ and minimize $\Gamma$, the entire
structure should be scaled down.  $F_{\text{min}} \propto
\beta^{1/2}$, while because of the additional contribution of the
cantilever's effective length $l_e \propto \beta$, $\tau_{\text{min}}
\propto \beta^{3/2}$.  Reducing all dimensions while preserving the
aspect ratio of a long and thin cantilever beam should thus optimize
its ultimate force and torque sensitivity. This necessity for further
miniturization has positioned bottom-up techniques as the fabrication
methods of the future.

In recent years, remarkable progress has been made in this direction
with force sensors made from doubly clamped
CNTs~\cite{Sazonovatunablecarbonnanotube2004}, suspended graphene
sheets~\cite{BunchElectromechanicalResonatorsGraphene2007}, and NW
cantilevers~\cite{GloppeBidimensionalnanooptomechanicstopological2014,RossiVectorialscanningforce2017,deLepinayuniversalultrasensitivevectorial2017a}. In
two separate papers, Moser et al. demonstrated the use of a CNT as a
sensitive force sensor with a thermally limited force sensitivity of
\SI{12}{\zepto\newton/\sqrt{\hertz}} at \SI{1.2}{\kelvin} in
2013~\cite{MoserUltrasensitiveforcedetection2013} and then of
\SI{1}{\zepto\newton/\sqrt{\hertz}} at \SI{44}{\milli\kelvin} in
2014~\cite{MoserNanotubemechanicalresonators2014}. Given their geometry, graphene
resonators are extremely difficult to apply in scanning probe
applications. Singly clamped NWs and CNTs on the other hand, when
arranged in the pendulum geometry -- that is, with their long axes
perpendicular to the sample surface -- are well-suited as scanning
probes. Their orientation prevents the tip from snapping into
contact~\cite{GysinLowtemperatureultrahigh2011}. When brought close to a surface, NWs
experience extremely low non-contact
friction~\cite{NicholNanomechanicaldetectionnuclear2012}, making near-surface
($<\SI{100}{\nano\meter}$) force sensitivities around
\SI{1}{\atto\newton/\sqrt{\hertz}}.  As a result, NWs have already
been used as scanning probes in a variety of proof-of-principle
experiments discussed in \cref{Sec:forces}.  

\subsection{Displacement detection}
\label{Sec:forcesesensing_displacement}
\begin{figure*}[t]
	\includegraphics[width=0.95\textwidth]{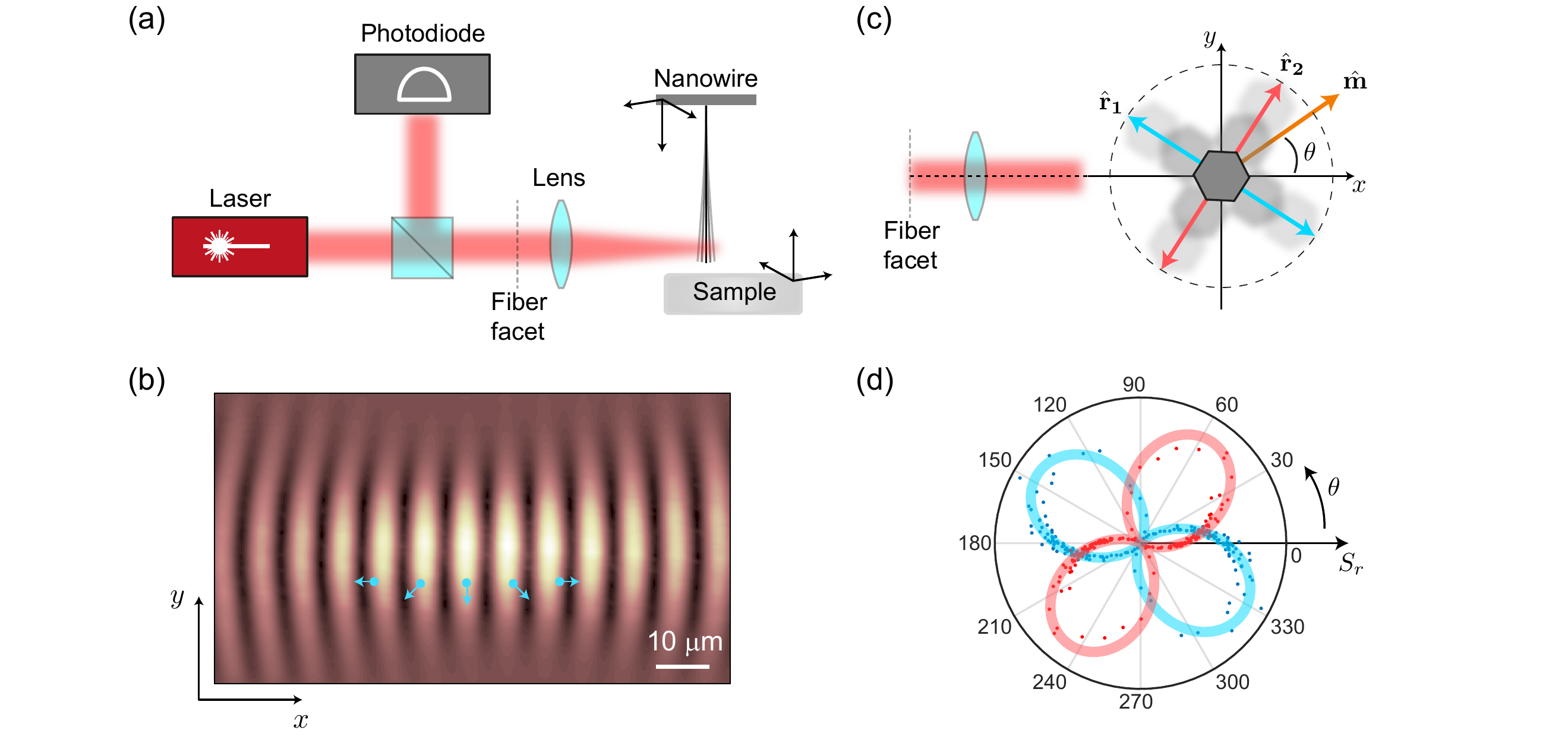}
	\caption{(a) Diagram of NW displacement detection setup. (b) Reflected intensity measured as a function of xy postion of NW tip showing the interference fringes resulting from the low-finesse interferometer in the detection path. Arrows indicate the direction of the gradient of the intensity for different positions, determining the direction of the measurement vector. (c) Diagram illustrating mode directions and measurement vector. (d) NW displacement PSD as a function of measurement angle.\vspace{1em}}
	\label{fig:detection}
\end{figure*}
As discussed above, the force tranducer sets the fundamental
sensitivity limits of any force sensor.  However, for a force to be
measured, the transducer's displacement must be detected.  Therefore,
in order to achieve the sensor's ultimate sensitivity, a displacement
detection scheme sensitive enough to detect the transducers thermal
motion is required.  The displacement of conventional AFM cantilevers
is measured through the deflection of a laser beam reflected off of
the cantilever and onto a split photodetector.  Although this scheme
works well for cantilevers with flat surfaces large enough to
specularly reflect a laser beam, it is not well-suited for NWs.  Given
their shape and sub-wavelength diameter, diffraction effects emerge
and an alternative means of displacement detection must be employed.

To circumvent this challenge, researchers have developed a number of
optical~\cite{SekaricNanomechanicalresonantstructures2002,AzakNanomechanicaldisplacementdetection2007,LiBottomupassemblylargearea2008,BelovMechanicalresonanceclamped2008,FaveroFluctuatingnanomechanicalsystem2009,RamosOptomechanicsSiliconNanowires2012}
and non-optical techniques, including electron
beam~\cite{TannerHighQGaNnanowire2007},
piezoresistive~\cite{HeSelfTransducingSiliconNanowire2008},
magnetomotive~\cite{FengVeryHighFrequency2007}, and
capacitive~\cite{TruittEfficientSensitiveCapacitive2007,Sazonovatunablecarbonnanotube2004}
displacement detection.  None of these, however, are compatible with
sensitive scanning probe microscopy (SPM) applications, in which the
NW cantilever’s thermal motion must be non-invasively detected while
its tip is positioned close to the surface of a sample.  Most
recently, a pair of optical techniques based on optical interferometry
or scattering have been shown to be compatible with SPM applications,
while at the same time providing access to both the amplitude and
direction of the NW
displacement~\cite{NicholDisplacementdetectionsilicon2008,GloppeBidimensionalnanooptomechanicstopological2014,RossiVectorialscanningforce2017,deLepinayuniversalultrasensitivevectorial2017a,RossiMagneticforcesensing2018a}.
A technique in which the electron beam in a scanning electron
microscope (SEM) is focused on a cantilever and the inelastically
scattered electrons are detected also resolved thermal displacement
fluctuations of a singly clamped CNT~\cite{TsioutsiosRealTimeMeasurementNanotube2017}.
This scheme was subsequently integrated into a SPM with the CNT as the
force transducer~\cite{SiriaElectronbeamdetection2017}.  Due to its flexibity and
relatively simple hardware required for its implementation, we will
focus here on the optical techniques.

Using Mie scattering analysis, it can be shown that a NW positioned
near the focus of an incident laser beam scatters sufficient light
such that a measurement of optical transmission is sufficient to
sensitively detect its displacement~\cite{SaniiHighSensitivityDeflection2010}.  The
technique uses the strong scattering intensity gradient as a
function of the NW position in the plane perpendicular to its axis
($xy$-plane), to convert NW displacement into a change in optical
transmission intensity $I_T$.  The gradient of this intensity as a
function of the NW position yields both the detection efficiency
$\epsilon = \left | \nabla I_T (x, y) \right |$ in $\text{W}/\text{m}$
and the direction along which displacement is detected $\hat{\mathbf{m}} =
\nabla I_T (x, y) / \left | \nabla I_T (x, y) \right |$
~\cite{GloppeBidimensionalnanooptomechanicstopological2014}.  The measured displacement is
therefore a projection of the dispacement of the two flexural modes
$m(t) = \hat{\mathbf{m}} \cdot (\mathbf{r_1}(t)+\mathbf{r_2}(t))$ along a
direction that depends on the position of the NW in the beam waist.
Using a split-photodiode detector, different response profiles can be
obtained from the sum and difference of the two sensors.  As a result,
displacement signals along two perpendicular directions can be
measured at the same time~\cite{deLepinayuniversalultrasensitivevectorial2017a}.  In this
way, even with modest optical powers, the thermal displacement of the
NW modes can be measured in two-dimensions.  The detection can be
used to fully characterize the response of each set of doublet
flexural modes to external forces and force gradients.  Ultimately,
this vectorial information allows for the reconstruction of the full
two-dimensional force field acting on the NW.  In addition, the
relative orientation of the doublet, i.e.\ whether these modes
oscillate orthogonal to each other or not, carries information about
whether the force field in which the NW is immersed is conservative or
not~\cite{LepinayEigenmodeorthogonalitybreaking2018}.

Another version of this detection scheme uses the interference between
light scattered back from the NW and light reflected from the cleaved
end of an optical fiber to measure NW displacement~\cite{RossiMagneticforcesensing2018a,FuDeterminingdirectionnanowire2017}.  In this case, a
fiber-based confocal reflection microscope is used to collect the
light scattered from the NW, as shown in Fig.~\ref{fig:detection}~(a).
The fiber's end and the NW form a low-finesse Fabry-Perot cavity
interferometer~\cite{NicholDisplacementdetectionsilicon2008,RossiVectorialscanningforce2017},
whose interferometric response depends on the position of the NW in
the beam waist, as seen in Fig.~\ref{fig:detection}~(b).  A fast
photo-receiver monitors variations in the reflected intensity $I_R$,
allowing for the sensitive detection of NW motion.  As before, the
efficiency $\epsilon = \left | \nabla I_R (x, y) \right |$ and the
direction of the detected displacement $\hat{\mathbf{m}} = \nabla I_R (x, y) /
\left | \nabla I_R (x, y) \right |$ depend on the variations of $I_R$
as a function of the NW position in the optical
waist~\cite{RossiMagneticforcesensing2018a}.  In this case, along the cavity's
axis, $\nabla I_R$ depends on an interference effect and therefore can
have a very high $\epsilon$, which can be improved by increasing the
cavity finesse.  Detection efficiency perpendicular to the cavity
axis, on the other hand, depends on the narrowness of the optical waist.
Using this method, the axis along which displacement is measured can
be changed either by moving the NW within the optical waist or by
tuning the laser excitation wavelength, which alters the interference
pattern shown in Fig.~\ref{fig:detection}~(b).  As with the scattering
based technique, this two-dimensional displacement detection technique
allows for both angular and spectral tomography of a NW's flexural
modes.  As an example, the displacement power spectral density of a
NW's thermally-excited fundamental mode doublet and its displacement
power density as a function of the angle in the $xy$-plane are shown in
Fig.~\ref{fig:detection}~(d).

Both the optical transmission and the interferometric techniques have
been successfully integrated into a SPM using a single NW as the force
transducer.  Although technical restrictions exist, such that there is
sufficient optical access to the NW (e.g.\ on the length of NW and the
distance of the scanning region from the sample edge), setups can be
designed to work with most samples of interest.

\section{Force microscopy with nanowire cantilevers}
\label{Sec:forces}
Recently, several proof-of-concept experiments have been performed that demonstrate the potential of force microscopy with NW cantilevers. First experiments focused on developing the optical detection of NW displacement\cite{NicholDisplacementdetectionsilicon2008,RamosOptomechanicsSiliconNanowires2012,RamosSiliconnanowireswhere2013} with high enough precision to observe the NW's thermomechanical noise, as well as on characterizing the mechanical properties of NW cantilevers\cite{NicholControllingnonlinearitysilicon2009,BraakmanNonlinearmotionmechanical2014,CadedduTimeResolvedNonlinearCoupling2016}. The promise of ultralow dissipation motivated researchers to implement NW cantilevers into scanning probe setups, for the sensitive detection of various types of forces. In a series of experiments\cite{NicholNanomechanicaldetectionnuclear2012,NicholNanoscaleFourierTransformMagnetic2013,NicholControllingnonlinearitysilicon2009}, Nichol et al. first used Si NWs as transducers in magnetic resonance force microscopy, exploiting their high mechanical frequencies and low dissipation to improve the sensitivity and resolution in measurements of small numbers of nuclear spins.

As mentioned before, when used in the pendulum geometry, NW cantilevers enable scanning probe microscopy of lateral forces. Such a mode of operation is of great interest for the detection of for instance frictional forces\cite{KisielSuppressionelectronicfriction2011} and for the detection of short-range non-central forces\cite{WeymouthNoncontactlateralforce2017}. Although one-dimensional dynamic lateral force microscopy can be realized using the torsional mode of conventional AFM cantilevers~\cite{PfeifferLateralforcemeasurementsdynamic2002,GiessiblFrictiontracedsingle2002,KawaiDynamiclateralforce2005,KawaiDirectmappinglateral2009,KawaiUltrasensitivedetectionlateral2010}, the ability to simultaneously image all vectorial components of nanoscale force fields is of great interest. Not only does it provide more information on tip-sample interactions, but it also enables the investigation of inherently 2D effects, such as the anisotropy or non-conservative character of specific interaction forces. This vectorial force sensing has now been demonstrated for a variety of types of interaction. 

The first implementation of a vectorial NW force sensor was realized by Gloppe et al.\cite{GloppeBidimensionalnanooptomechanicstopological2014} in an experiment where the optical force field of a laser beam focused onto a NW was mapped out. Optomechanical interactions lie at the basis of a very fruitful line of research in which quantum states of motion are studied using optical fields, for instance in optical resonators. To fully describe such optomechanical interactions, one needs to take into account their non-conservative nature, and their spatial mapping necessitates the use of a fully vectorial force sensor as described by Gloppe et al. Interestingly, this non-conservative character of the optomechanical interaction can induce strong coupling between orthogonal mechanical modes. Interestingly, the vectorial nature of the optomechanical interaction can induce strong coupling between orthogonal mechanical modes, which can lead to non-trivial forms of back-action on the NW motion.

In a next step, vectorial NW force sensors were implemented in scanning probe setups, in which the NW cantilever could be scanned over a sample surface\cite{RossiVectorialscanningforce2017, deLepinayuniversalultrasensitivevectorial2017a}. These setups allowed for the measurement of two-dimensional maps of the sample surface, in which forces arising from Coulombic interactions, chemical bonding, and van der Waals interactions provide a topographic contrast. The NW cantilevers complement conventional AFM in providing a full determination of the four lateral force derivatives. Together, these experiments show the potential of NW cantilevers as ultrasensitive bidimensional force transducers and provide experimental quantification of sensitivities and resolution. Force sensitivities were measured to be in the $\textrm{aN Hz}^{-1/2}$ range. This is on par with or slightly better than sensitivities reached with conventional AFM. However, there is great potential for improvement of the force sensitivity, through optimization of NW geometry and surface properties. Spatial resolution was limited, as expected, by the NW tip diameter, which for these first experiments ranged between 100 and 350\,nm. Also here, there is a clear route for improvement, since several types of NWs and nanotubes with tip diameters that are 10-100 times smaller have been grown.

A further exciting new development is the use of NW cantilevers with magnetic tips for sensitive and vectorial detection of magnetic forces. In recent years, there has been a flurry of activity in developing
nanometer-scale magnetic imaging technology. These efforts are driven by a number outstanding questions in spintronics -- such as how to control magnetic skyrmions -- and in mesoscopic transport -- such as
how current flows in topological insulators and two-dimensional materials. Scanning probe microscopy, in particular, has made remarkable improvements in both the sensitivity and resolution of magnetic imaging.  Some of the most successful tools are magnetic force microscopy (MFM), spin-polarized scanning tunneling microscopy, as well as scanning magnetometers based on nitrogen-vacancy centers in diamond, Hall-bars, and superconducting quantum interference devices (SQUIDs). Despite this progress, it is now becoming clear that
nanomechanical sensors, in general, and NWs in particular, provide a huge untapped opportunity in magnetic sensing. First results of NW MFM were obtained using a NW with a magnetic segment grown at its tip\cite{RossiMagneticforcesensing2018a}.


\subsection{Optical force sensing}
\label{Sec:forces_optical}
\begin{figure}[t]
	\includegraphics[width=0.45\textwidth]{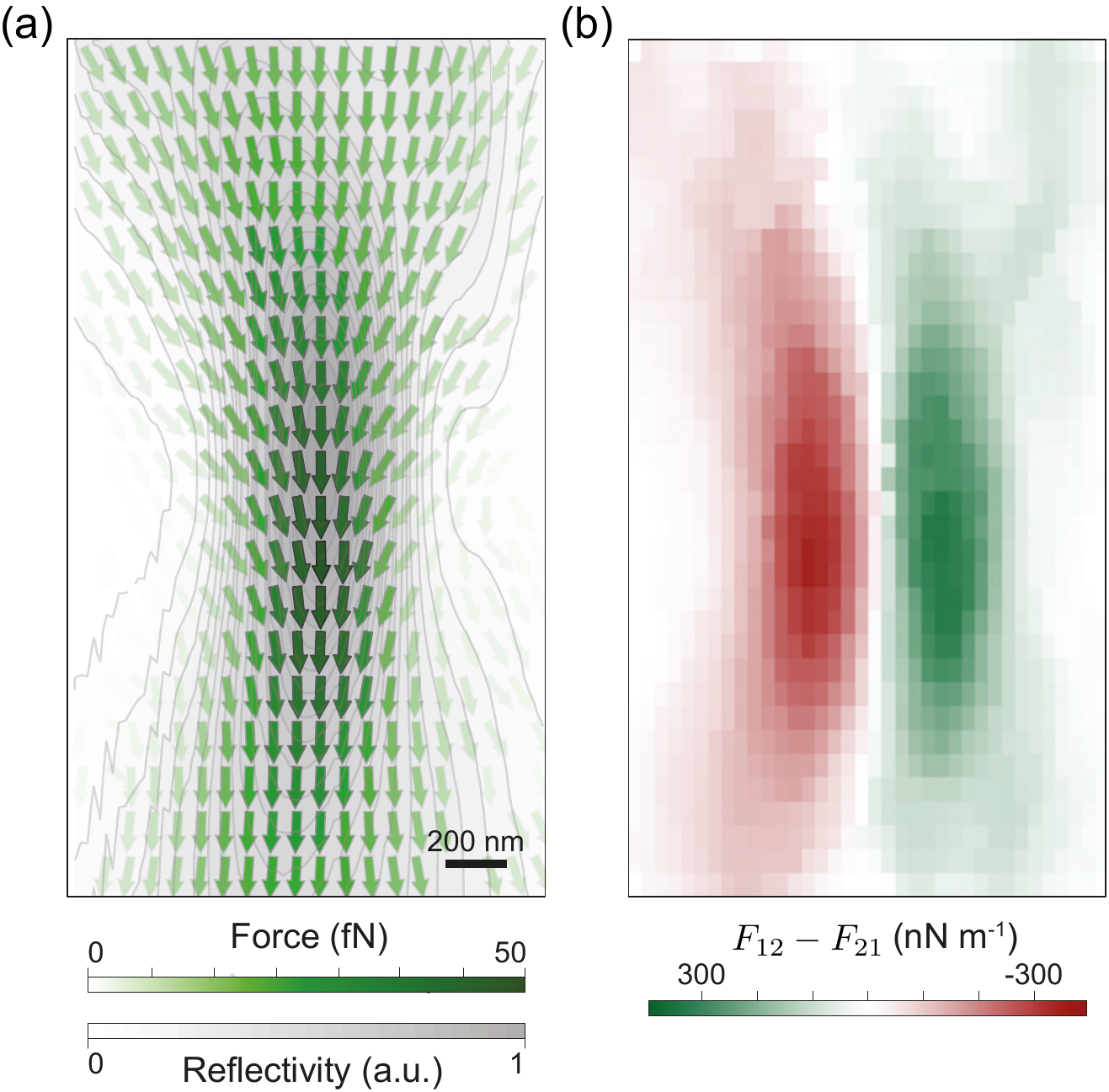}
	\caption{(a) Measured spatial map of optomechanical force field on NW tip placed inside focused laser beam. (b) Spatial map of the local curl derived from measurement displayed in (a). Adapted from Ref.\cite{GloppeBidimensionalnanooptomechanicstopological2014}.\vspace{1em}}
	\label{fig:optical}
\end{figure}

Optical setups used to detect the displacement of a cantilever, be it based on interferometry or scattering of light, can also be applied to study optomechanical interactions. The NW cantilever with its two orthogonal flexural modes gives the opportunity to investigate the vectorial nature of optomechanical coupling. Gloppe et al. have performed such a study, by placing a NW cantilever inside the optical field of a laser beam tightly focused by a microscope objective with a high numerical aperture\cite{GloppeBidimensionalnanooptomechanicstopological2014}. In this experiment, the optical intensity of the field was modulated with a frequency that was swept through both orthogonal mode resonant frequencies, providing driving forces through the optomechanical interaction. A second, much weaker laser beam was used to probe the thermal force noise of the NW, allowing the determination of the direction of the NW modes following the method outlined in \cref{Sec:mechanics_twomodes}. The method of \cref{fig:forces} was then used to acquire a map of the magnitude and direction of the in-plane optomechanical force. The force field of \cref{fig:optical}a shows a converging and diverging vector flow corresponding to the waist area of the focused laser beam. The optomechanical interaction is in general non-conservative and therefore the force field acting on the NW can possess a non-zero curl. The curl can be found by differentiating the vector force field and was determined to be maximum at the waist, on each side of the optical axis (see \cref{fig:optical}b). Furthermore, by measuring the delay between the laser intensity modulation and the NW response it was shown that the optomechanical force was in phase with the intensity modulation, as expected for a dissipative optomechanical coupling. The authors report a force sensitivity in the $\textrm{aN Hz}^{-1/2}$ range at room temperature.
\subsection{Electrical force sensing}
\label{Sec:forces_electrical}
\begin{figure*}[t]
	\includegraphics[width=0.95\textwidth]{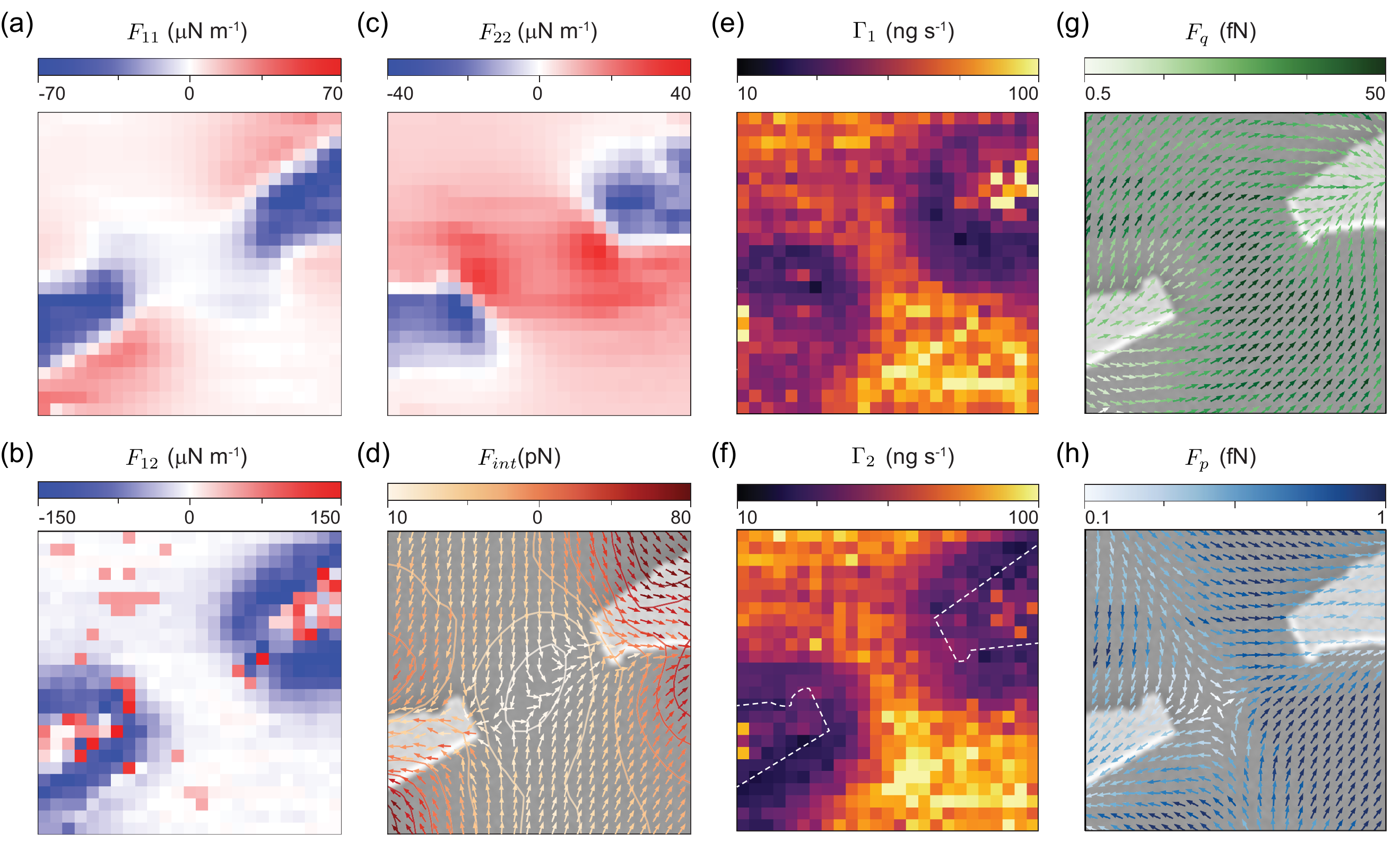}
	\caption{(a-c) Measurements of $F_{11}$, $F_{12}$, and $F_{22}$ tip-sample force derivatives as a function of NW tip position over patterned sample surface. The scale bar corresponds to \SI{1}{\micro\meter} (d) Vector plot of in-plane force field derived by numerically integrating the derivatives of (a-c). The plot is overlaid on a scanning electron micrograph of the sample with patterned gate electrodes. (e), (f) Dissipation of each orthogonal mode as a function of NW tip position. (g) Vector plot of $\textbf{F}_q$ induced by gate electric field on the charged NW. (h) Vector plot of $\textbf{F}_p$ induced by gate electric field on the polarizable NW.
	Adapted from Ref.\cite{RossiVectorialscanningforce2017}.\vspace{1em}}
	\label{fig:efield}
\end{figure*}
In scanning probe measurements of a sample surface, such as those made in conventional atomic force microscopy, the underlying sample-cantilever forces are typically of electrical origin. These forces include contributions from van der Waals forces and various other interactions of electrostatic origin. The measured forces closely follow the geometrical pattern of the sample surface, enabling the reconstruction of sample topography. In contrast to conventional AFM, NW cantilevers arranged in the pendulum geometry allow for the measurement of vectorial topographical maps of sample surfaces through the measurement of in-plane forces and force derivatives. In two experiments\cite{RossiVectorialscanningforce2017,deLepinayuniversalultrasensitivevectorial2017a}, researchers recently demonstrated the use of NW cantilevers to measure such topographical maps.\\

\cref{fig:efield} shows various measurements involving a specific sample featuring \SI{200}{\nano\meter} thick Au gate electrodes lithographically defined on a planar Si substrate. By scanning a NW cantilever over the sample surface, vectorial maps of the tip-sample forces and forces derivatives, as well as of the dissipation of both NW modes, were measured\cite{RossiVectorialscanningforce2017}. \cref{fig:efield}a-c show maps of in-plane spatial force derivatives $F_{11}$, $F_{22}$, and $F_{12}$ (since electrical potentials are conservative, here $F_{21} = F_{12}$), through measurements of the frequency shift and oscillation direction of the two orthogonal flexural modes of the NW. The measured in-plane force derivatives can be integrated to produce a map of in-plane tip-sample forces, up to an integration constant corresponding to a constant force in the plane. \cref{fig:efield}d shows a map of the force field $F_{int}$ extracted from such an integration and confirms that the measured forces are roughly perpendicular to the edges defined by the sample topography. Furthermore, \cref{fig:efield}e, f show maps of measured dissipations $\Gamma_i$, extracted from the mode linewidths. Dissipation was measured to be nearly isotropic in the plane and was affected mostly by the different materials and tip-sample spacing over electrodes and substrate. In principle however, the demonstrated ability to produce vectorial maps of dissipation moreover that NW cantilevers could find application in the study of anisotropic non-contact friction, which is important for instance for the study of superlubricity\cite{VilhenaSlipperyeverydirection2018}.

In a next step, the authors generated an alternating electric field $\textbf{E}(\textbf{r},t)$ using the patterned electrodes to drive the motion of both NW modes. In this way, vectorial in-plane maps of the driving forces were obtained (see \cref{fig:efield}g, h), with a thermally limited sensitivity of 5\,$\textrm{aN Hz}^{-1/2}$ at a temperature of \SI{4}{\kelvin}. Two types of driving forces resulted from $\textbf{E}(\textbf{r},t)$: $\textbf{F}_q = q\textbf{E}$ and $\textbf{F}_p = -\nabla(\alpha|\textbf{E}|^2)$. Here $q$ is the net charge on the NW tip and $\alpha$ the polarizability of the NW. Since $\textbf{F}_q$ shows a linear dependence on the magnitude of the electric field, while that of $\textbf{F}_p$ is quadratic, these two forces are spectrally separated and could therefore be distinguished in measurement. Interestingly, such force measurements also allow the characterization of the NW itself. By comparing the magnitude of the measured forces with that of the applied electric field, the average values of the constants $q$ and $\alpha$ could in this case be determined.

In a similar setting, Mercier de L\'epinay et al. measured vectorial force fields and force field derivatives arising from the electrostatic interaction of a voltage-biased metallic sample of pyramidical shape with a NW cantilever\cite{deLepinayuniversalultrasensitivevectorial2017a}. The authors furthermore reconstructed a three-dimensional image of the force field derivatives by scanning the NW vertically away from the sample, up to a tip-sample distance of several microns.

\subsection{Magnetic force microscopy}
\label{Sec:forces_MFM}

In recent years, there has been a flurry of activity in developing
nanometer-scale magnetic imaging technology.  These efforts are driven
by a number outstanding questions in spintronics -- such as how to
control magnetic skyrmions -- and in mesoscopic transport -- such as
how current flows in topological insulators and two-dimensional
materials.  Scanning probe microscopy, in particular, has made
remarkable improvements in both the sensitivity and resolution of
magnetic imaging.  Some of the most successful tools are magnetic
force microscopy (MFM), spin-polarized scanning tunneling microscopy,
as well as scanning magnetometers based on nitrogen-vacancy centers in
diamond, Hall-bars, and superconducting quantum interference devices
(SQUIDs).  Despite this progress, it is now becoming clear that
nanomechanical sensors, in general, and NWs in particular, provide a
huge untapped opportunity in magnetic sensing.

The first magnetic SPM, MFM, was introduced in the late 1980s as a
natural extension of AFM. These days, it is performed in air, liquid,
vacuum, and at a variety of temperatures.  Under ideal conditions,
state-of-the-art MFM can reach spatial resolutions down to
10~nm~\cite{SchmidExchangeBiasDomain2010}, though more typically around 100
nm.  In 2009, its application to magnetic resonance, magnetic
resonance force microscopy (MRFM), resulted in the first demonstration
of three-dimensional nuclear magnetic resonance imaging (MRI) with
nanometer-scale resolution~\cite{DegenNanoscalemagneticresonance2009}.

NW scanning force sensors with proper functionalization of their tips,
can be used to measure weak magnetic forces.  Recent experiments have
shown that a NW's high force sensitivity -- when combined with a
highly concentrated and strongly magnetized tip -- gives it an
exquisitive sensitivity to magnetic field
gradients~\cite{RossiMagneticforcesensing2018a}.  Using a MnAs-tipped GaAs NW,
the authors demonstrated a sensitivity to $\SI{11}{\milli \tesla /
	\meter \sqrt{\hertz}}$.  Having quantified the NW's response to
magnetic field gradients, the authors calculate its sensitivity to
other magnetic field sources, including a magnetic moment (dipole
field), a superconducting vortex (monopole field), or an infinitely
long and thin line of current~\cite{KirtleyFundamentalstudiessuperconductors2010}.  In
particular, they expect a moment sensitivity of 54~$\mu_B
/\sqrt{\text{Hz}}$, a flux sensitivity of 1.3~$\mu \Phi_0
/\sqrt{\text{Hz}}$, and line-current sensitivity of $9\;\text{nA} /
\sqrt{\text{Hz}}$ at a tip-sample spacing of \SI{250}{\nano\meter}.  Such sensitivities compare favorably to those of
other magnetic microscopies, including scanning Hall microscopy,
magneto-optic microscopy, scanning SQUID microscopy, and scanning
nitrogen-vacancy magnetometry.  Furthermore, magnet-tipped NWs have a
huge potential for improvement as probes of weak magnetic field
patterns if tips sizes and tip-sample spacings can be further reduced.
\begin{figure*}[t]
	\includegraphics[width=0.95\textwidth]{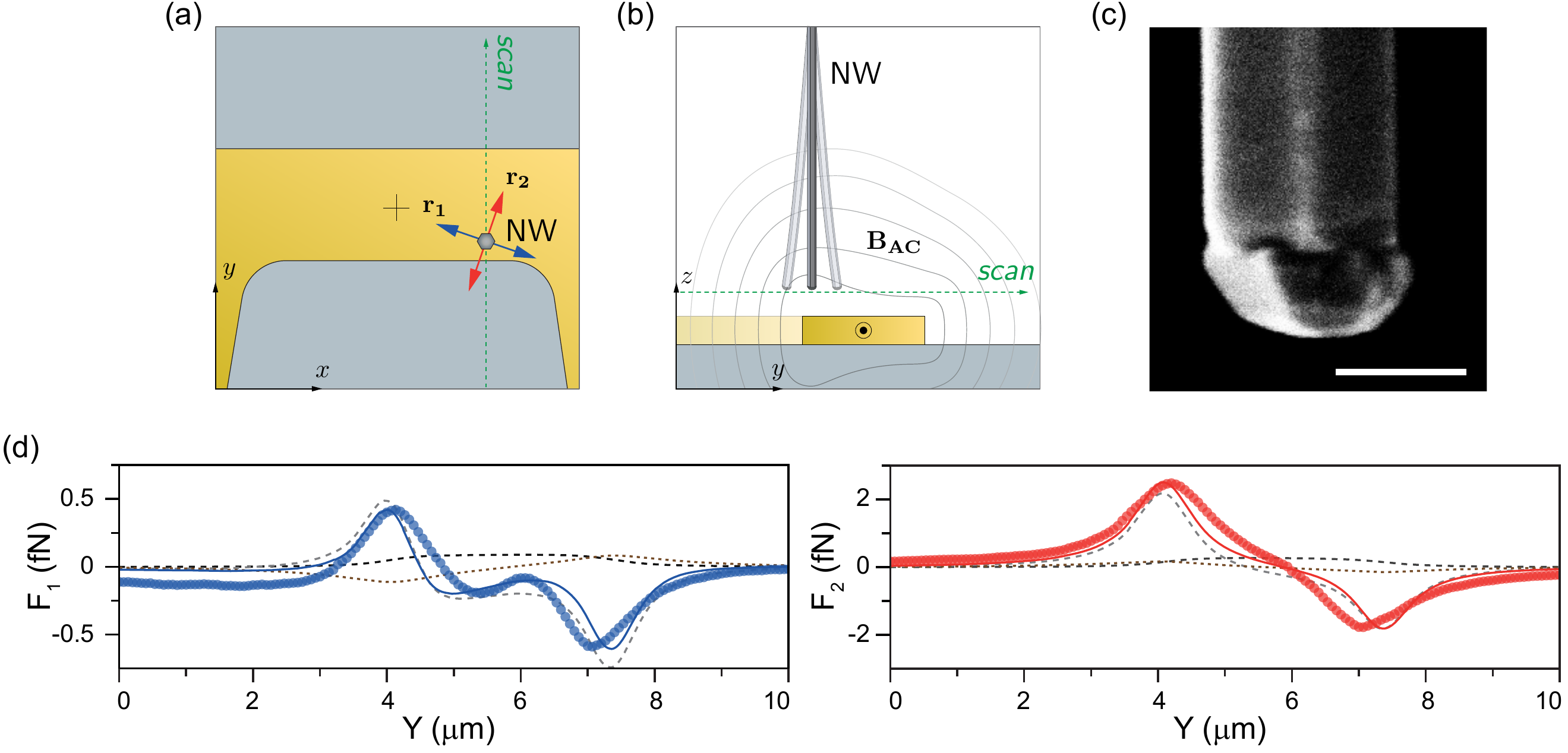}
	\caption{(a), (b) Schematic illustration of MFM experiment with NW cantilever and current-carrying Au wire. The NW scan direction used to produced measurements in (c) is indicated by the green line in top view (a) and side view (b). $B_{AC}$ corresponds to the magnetic field produced by the alternating current in the wire, $B$ to an externally applied magnetic field.(c) Scanning electron micrograph of the tip of a GaAs NW featuring a MnAs segment at the end. The scalebar corresponds to \SI{100}{\nano\meter}. (d) Plots of the measured (dotted line) and calculated (solid line) forces driving the first (blue) and the second (red) mode over the line scan indicated in (a) and (b). For each plot three distinct drive contributions are shown as dashed lines: the monopole (black), dipole (gray) and torque (magenta) terms. Adapted from Ref.\cite{RossiMagneticforcesensing2018a}.\vspace{1em}}
	\label{fig:MFM}
\end{figure*}

In addition to improved sensitivity, NW MFM provides other potential
advantages compared to conventional MFM.  First, scanning in the
pendulum geometry with the NW oscillating in the plane of the sample
has the characteristics of lateral MFM.  This technique, which is
realized with the torsional mode of a conventional cantilever,
distinguishes itself from the more commonly used tapping-mode MFM in
its ability to produce magnetic images devoid of spurious
topography-related contrast and in a demonstrated improvement in
lateral spatial resolution of up to
15\%~\cite{KaidatzisTorsionalresonancemode2013}.  Second, the nanometer-scale
magnetic particle at the apex of the NW force sensor minimizes the
size of the MFM tip, allowing for optimal spatial resolution and
minimal perturbation of the investigated sample.

There have been a number of efforts at creating high-resolution MFM
tips on conventional cantilevers, including attaching coated CNT
tips~\cite{DengMetalcoatedcarbonnanotube2004}, milling tips by
FIB~\cite{PhillipsHighresolutionmagnetic2002}, or using electron beam induced
deposition
techniques~\cite{LauPropertiesapplicationscobaltbased2002,KoblischkaImprovementslateralresolution2003}.
Despite this work, non-invasive, high-resolution MFM tips are still
needed.  In particular, the tips with the smallest volumes realized
thus far, tend to have rather undefined magnetization directions due
to amorphous magnetic materials and nearly symmetric particle shape.
Recent results show that small single crystal magnetic particles with
defined and predictable magnetization directions can be realized atop
a single NW~\cite{HubmannEpitaxialGrowthRoomTemperature2016,RossiMagneticforcesensing2018a}.  In
particular, Rossi et al.\ carried out measurements of dynamic
cantilever magnetometry~\cite{RosselActivemicroleversminiature1996,HarrisIntegratedmicromechanicalcantilever1999, StipeMagneticDissipationFluctuations2001} to extract the magnetic
properties of each MnAs tip from the mechanical response of the NW to
a uniform external magnetic field.  These measurements, along with
measurements of the NW's response to a well-known magnetic field
profile generated by a lithographically patterned wire, confirmed that
in remanence the magnetic tips are -- in most cases -- strongly
magnetized and dipole-like. \cref{fig:MFM}a and b show the experimental setup in which the magnet-tipped NW is scanned across a current-carrying wire.  An SEM of the MnAs magnetic tip is shown in Fig. 9 c. The magnetic force measured along the two mode directions is shown in Fig. 9 d, showing a response consistent with a remanent dipole-like magnetic tip.  These results bode well for the
resolution possible with NW MFM probes.  In principle, as discussed by
van Schendel et al., detailed analysis of MFM tips shows that the
closer a tip approaches an ideal magnetic dipole, the better its
sensitivity to high spatial frequencies, and therefore the higher its
potential resolution~\cite{Schendelmethodcalibrationmagnetic2000}.

The prospect of increased sensitivity and resolution, combined with
few restrictions on operating temperature, make NW MFM ideally suited
to investigate nanometer-scale spin textures, skyrmions,
superconducting and magnetic vortices, as well as ensembles of
electronic or nuclear spins.  Non-invasive magnetic tips may also open
opportunities to study current flow in 2D materials and topological
insulators.  The ability of a NW sensor to map all in-plane spatial
force
derivatives~\cite{RossiVectorialscanningforce2017,deLepinayuniversalultrasensitivevectorial2017a}
should provide fine detail of stray field profiles above magnetic and
current carrying samples, in turn providing detailed information on
the underlying phenomena.  Directional measurements of dissipation may
also prove useful for visualizing domain walls and other regions of
inhomogeneous magnetization.  As shown by Gr\"utter et al.,
dissipation contrast, which maps the energy transfer between the tip
and the sample, strongly depends on the sample's nanometer-scale
magnetic structure~\cite{GrutterMagneticdissipationforce1997}.

\subsection{Magnetic resonance force microscopy}

So far, the most developed application for magnetic force sensing with
NWs has been MRFM for nanometer-scale magnetic resonance imaging
(nano-MRI).  The method relies on an ultra-sensitive mechanical force
sensor to detect the magnetic resonance from tiny ensembles of nuclear
spins.

Conventional MRI techniques employ pick-up coils to detect the small
changes in magnetic field induced by flipping nuclear moments
contained in a sample. These magnetic signals are so weak that
conventional instruments cannot resolve objects smaller than several
micrometers -- about the size of a small cell~\cite{Ciobanu3DMRmicroscopy2002}.
MRFM improves on this sensitivity by mechanically detecting the
magnetic forces produced by nuclear moments.  A mechanical resonator
is used to sense the forces arising between nuclear moments in a
sample and a nearby source of magnetic field gradient.  In MRFM, RF
pulses cause nuclear moments in a sample to periodically flip,
generating an oscillating force on the mechanical resonator.  These
alternating forces, in turn, drive the resonator to oscillate.
Force-detected MRI is more sensitive to nanometer-scale samples than
conventional techniques because much smaller detectors can be
made. For the inductive technique to be sensitive, the size of the
pick-up coil must be similar to the size of the sample.  For
nanometer-scale samples, this is practically impossible.  On the other
hand, high-quality cantilevers can have dimensions far below a
micrometer such that the sample mass is significant compared to the
bare resonator mass.  Such mechanical transducers enabled the current
state-of-the-art force-detected nano-MRI, which was demonstrated by
Degen et al. in 2009~\cite{DegenNanoscalemagneticresonance2009}.  In this work,
researchers captured 3D images of individual tobacco mosaic virus
(TMV) particles with a resolution better than 10 nm along each
dimension.  The technique has the unique capability to image the
interior of nanometer-scale objects non-invasively and with intrinsic
chemical selectivity.  Despite a number of further refinements and
demonstrations~\cite{PoggioForcedetectednuclearmagnetic2010,PoggioForceDetectedNuclearMagnetic2018},
improvements in MRFM sensitivity and resolution have stalled in recent
years, leaving a number of technical obstacles to be overcome for the
technique to become a useful tool for biologists and materials
scientists.

Foremost among these obstacles is the reduction in the mechanical
dissipation of the cantilever sensor.  Lower mechanical dissipation
would yield better force sensitivities and therefore sensitivity to
smaller numbers of nuclear spins.  Such an improvement would result in
nano-MRI with improved resolution.  In the last few years, the
development and application of NW cantilevers to MRFM has been making
promising steps in this direction.  In 2012, Nichol et al.\ used a Si
NW force transducer in an MRFM experiment detecting $^1$H in a
nanometer-scale polystyrene sample~\cite{NicholNanomechanicaldetectionnuclear2012}.
During the measurements they achieved a thermally limited force
sensitivity of around $\SI{1}{\atto\newton / \sqrt{\hertz}}$ at a
spacing of \SI{80}{\nano\meter} from the surface at \SI{8}{\kelvin},
which is significantly lower than was measured at
\SI{300}{\milli\kelvin} in the TMV
experiment~\cite{DegenNanoscalemagneticresonance2009}.  This improvement is largely
due to the ultra-low native dissipation of the NWs in comparison to
top-down ultrasensitive cantilever and to their drastically reduced
surface dissipation.  In fact, Nichol et al., show that at a
tip-surface spacing of \SI{7}{\nano\meter}, a typical Si NW
experiences nearly a factor of 80 less surface dissipation and factor
of 250 less total dissipation than audio frequency cantilevers under
similar conditions.  The mechanisms behind this difference are not
completely clear; the small cross-sectional area of a NW may decreases
its coupling to the surface or, perhaps, the spectral density of
surface fluctuations is lower at the MHz resonant frequencies of the
NWs that at the kHz resonant frequencies of the cantilevers.

This ground-breaking work established NW oscillators as ultrasensitive
cantilevers for MRFM detection.  The measurement protocol that was
developed for the NW transducers uses a nanoscale current-carrying
wire to generate both time-dependent RF magnetic fields and
time-dependent magnetic field gradients.  This protocol, known as
MAGGIC, may ultimately open new avenues for nanoscale magnetic
resonance imaging with more favorable SNR
properties~\cite{NicholNanoscaleFourierTransformMagnetic2013}.

Given that nanometer-scale MRFM requires intense static magnetic field
gradients, both NMR spectroscopy and uniform spin manipulation using
RF pulses have always been difficult to implement in such
measurements.  As a result, MRFM experiments often rely on inherently
slow adiabatic passage pulses, which limited the mechanical transducer
to resonance frequencies in the few kHz regime.  In addition,
conventional pulsed magnetic resonance techniques cannot be applied to
nanometer-scale MRFM because statistical spin fluctuations often
exceed the Boltzmann spin polarization~\cite{herzog_boundary_2014}.
In this regime, the projection of the sample magnetization along any
axis fluctuates randomly in time.

In their article, Nichol et al.\ presented a new paradigm in
force-detected magnetic resonance that overcomes both challenges to
enable pulsed nuclear magnetic resonance in nanometer-size
statistically polarized samples.  The first challenge was solved by
using the nanometer-scale constriction to generate both large RF
fields and large magnetic field gradients.  In this way, the authors
were able to turn their magnetic field gradients and on and off at
will.  The method allowed the use of high-frequency mechanical
resonances, such as those provided by a NW.  Using a scheme similar to
conventional MRI, switchable gradients in static and RF field encoded
the Fourier transform of the 2D spin density into their spin signal.
As a result, they were able to reconstruct a 2D projection of the
$^1$H density in a polystyrene sample with roughly 10-nm resolution.
The protocol was able to function in the statistically polarized
regime because the authors periodically applied RF pulses, which
create correlations in the statistical polarization of a solid organic
sample.  The spin-noise correlations were then read-out using gradient
pulses generated by ultra-high current densities in the nanoscale
metal constriction.  The authors also showed that Fourier-transform
imaging enhances sensitivity via the multiplex advantage for
high-resolution imaging of statistically polarized samples.  Most
importantly, the protocol established a method by which all other
pulsed magnetic resonance techniques can be used for nanoscale imaging
and spectroscopy.  Recent work by Rose et al., combines the high spin
sensitivity of NW-based magnetic resonance detection with
high-spectral-resolution NMR
spectroscopy~\cite{RoseHighResolutionNanoscaleSolidState2018}.  The authors make use
of the resulting enhancement in nuclear spin coherence times to
perform Fourier-transform imaging of proton spins with a
one-dimensional slice thickness below \SI{2}{\nano\meter}.

Given the potential for even more sensitive NW transducers, these
proof-of-concept experiments bode well for increasing nano-MRI
sensitivity and resolution.  Even without improvement in sensitivity,
the authors' technique could also be extended to enable full 3D
encoding with constrictions capable of producing two orthogonal static
gradients~\cite{NicholNanoscalemagneticresonance2014}.  More generally, the approach
serves as a model for applying sophisticated pulsed magnetic resonance
schemes from conventional MRI to the nanometer-scale version.

\section{Outlook}
\label{Sec:outlook}
The application of NW cantilevers as sensitive force transducers is in
the early stages.  Nevertheless, the
demand for more sensitive techniques has positioned nanometer-scale
mechanical structures -- and NWs in particular -- as the transducers
of the future.  At present, the promise of these devices has been
demonstrated in a small number of proof-of-concept experiments.  As
discussed in this review, these methods and applications demonstrate
the capabilities of NW transducers and their unique advantages,
especially in the detection of weak forces.  Nevertheless, much work
remains to be done before NWs become part of the standard tool-box of
scanning probe microscopies.

Force sensitivity can undoubtedly be improved by developing longer and
thinner NW transducers.  So far, researchers have not made concerted
efforts to optimize NW geometry, leaving much to be gained by such
work.  Better spatial resolutions could also be achieved by developing
NWs with sharper tips, through, for example, specialized growth
techniques or by focussed ion beam milling.  By carrying out SPM
experiments with such sharp NWs in ultra-high vacuum conditions,
atomic resolution should be possible.  Combined with its vectorial
force sensing capability, such a NW scanning probe could be used to
reveal the anisotropy of atomic bonding forces.

Higher force sensitivity would also translate into more sensitive
measurements of dissipation and non-contact friction.  AFM in the
pendulum geometry -- the same geometry used for NWs -- is ideally
suited to measurements of nanometer-scale dissipation.  Such
measurements have recently been used to detect superconducting~\cite{KisielSuppressionelectronicfriction2011} and bulk structural phase
transitions~\cite{KisielNoncontactAtomicForce2015}.  They can also shed light
on concepts such as superlubricity, atomic-scale friction, and quantum
friction~\cite{VolokitinQuantumFriction2011}.  Furthermore, energy
dissipation plays a central role in the breakdown of topological
protection, the loss of quantum information, and disorder-assisted hot
electrons scattering in graphene~\cite{HalbertalNanoscalethermalimaging2016}.  The
ability to map tip-sample force fields and energy losses makes NW
transducers ideal for investigating how and where energy leaks.

Finally, the combination of high force sensitivity, high spatial
resoltution, and low invasivement of magnetic NW probes has the
potential to expand the applicability of MFM.  Such sensors could
image the stray field thin ferromagnetic layers and other magnetic
nanostructures hosting non-uniform states too fragile for
conventional MFM.  For such applications, efforts must first focus on
producing smaller magnetic tips.  Alternatively, the production
of different types of magnet-tipped NWs could be attempted, through
the evaporation of magnetic caps on sharp non-magnetic NWs or by direct
focused ion beam induced deposition. By combining NW transducers with
ferromagnetic resonance techniques, the spatial imaging of spin waves and the dynamics of
individual skyrmions may also become possible.
%
%
%
%
%
%

\newpage
\bibliographystyle{iopart-num}
\bibliography{NWReview}

\providecommand{\newblock}{}
\begin{thebibliography}{100}
\expandafter\ifx\csname url\endcsname\relax
  \def\url#1{{\tt #1}}\fi
\expandafter\ifx\csname urlprefix\endcsname\relax\def\urlprefix{URL }\fi
\providecommand{\eprint}[2][]{\url{#2}}

\bibitem{GloppeBidimensionalnanooptomechanicstopological2014}
Gloppe A, Verlot P, {Dupont-Ferrier} E, Siria A, Poncharal P, Bachelier G,
  Vincent P and Arcizet O 2014 {\em Nature Nanotechnology\/} {\bf 9} 920--926
  ISSN 1748-3387

\bibitem{ThillosenStateStrainSingle2006}
Thillosen N, Sebald K, Hardtdegen H, Meijers R, Calarco R, Montanari S, Kaluza
  N, Gutowski J and L\"uth H 2006 {\em Nano Letters\/} {\bf 6} 704--708 ISSN
  1530-6984

\bibitem{GrutterMagneticdissipationforce1997}
Gr\"utter P, Liu Y, LeBlanc P and D\"urig U 1997 {\em Applied Physics
  Letters\/} {\bf 71} 279--281 ISSN 0003-6951

\bibitem{KisielSuppressionelectronicfriction2011}
Kisiel M, Gnecco E, Gysin U, Marot L, Rast S and Meyer E 2011 {\em Nature
  Materials\/} {\bf 10} 119--122 ISSN 1476-4660

\bibitem{CockinsExcitedStateSpectroscopyIndividual2012}
Cockins L, Miyahara Y, Bennett S~D, Clerk A~A and Grutter P 2012 {\em Nano
  Letters\/} {\bf 12} 709--713 ISSN 1530-6984

\bibitem{NicholNanomechanicaldetectionnuclear2012}
Nichol J~M, Hemesath E~R, Lauhon L~J and Budakian R 2012 {\em Physical Review
  B\/} {\bf 85} 054414

\bibitem{NicholNanoscaleFourierTransformMagnetic2013}
Nichol J~M, Naibert T~R, Hemesath E~R, Lauhon L~J and Budakian R 2013 {\em
  Physical Review X\/} {\bf 3} 031016

\bibitem{RoseHighResolutionNanoscaleSolidState2018}
Rose W, Haas H, Chen A~Q, Jeon N, Lauhon L~J, Cory D~G and Budakian R 2018 {\em
  Physical Review X\/} {\bf 8} 011030

\bibitem{RossiVectorialscanningforce2017}
Rossi N, Braakman F~R, Cadeddu D, Vasyukov D, T\"ut\"unc\"uoglu G, {Fontcuberta
  i Morral} A and Poggio M 2017 {\em Nature Nanotechnology\/} {\bf 12} 150--155
  ISSN 1748-3387

\bibitem{BraakmanCoherentTwoModeDynamics2018}
Braakman F~R, Rossi N, T\"ut\"unc\"uoglu G, i~Morral A~F and Poggio M 2018 {\em
  Physical Review Applied\/} {\bf 9} 054045

\bibitem{RossiMagneticforcesensing2018a}
Rossi N, Gross B, Dirnberger F, Bougeard D and Poggio M 2018

\bibitem{CadedduTimeResolvedNonlinearCoupling2016}
Cadeddu D, Braakman F~R, T\"ut\"unc\"uoglu G, Matteini F, R\"uffer D,
  {Fontcuberta i Morral} A and Poggio M 2016 {\em Nano Letters\/} {\bf 16}
  926--931 ISSN 1530-6984

\bibitem{MontinaroQuantumDotOptoMechanics2014}
Montinaro M, W\"ust G, Munsch M, Fontana Y, {Russo-Averchi} E, Heiss M,
  {Fontcuberta i Morral} A, Warburton R~J and Poggio M 2014 {\em Nano
  Letters\/} {\bf 14} 4454--4460 ISSN 1530-6984

\bibitem{FosterTuningNonlinearMechanical2016}
Foster A~P, Maguire J~K, Bradley J~P, Lyons T~P, Krysa A~B, Fox A~M, Skolnick
  M~S and Wilson L~R 2016 {\em Nano Letters\/} ISSN 1530-6984

\bibitem{PairisShotnoiselimited2018}
Pairis S, Donatini F, Hocevar M, Tumanov D, Vaish N, Claudon J, Poizat J~P and
  Verlot P 2018 {\em arXiv:1803.09312 [cond-mat]\/} (\textit{Preprint}
  \eprint{1803.09312})

\bibitem{deLepinayuniversalultrasensitivevectorial2017a}
{de L\'epinay} L~M, Pigeau B, Besga B, Vincent P, Poncharal P and Arcizet O
  2017 {\em Nature Nanotechnology\/} {\bf 12} 156--162 ISSN 1748-3387

\bibitem{LepinayEigenmodeorthogonalitybreaking2018}
de~L\'epinay L~M, Pigeau B, Besga B and Arcizet O 2018 {\em Nature
  Communications\/} {\bf 9} 1401 ISSN 2041-1723

\bibitem{PerisanuHighfactormechanical2007}
Perisanu S, Vincent P, Ayari A, Choueib M, Purcell S~T, Bechelany M and Cornu D
  2007 {\em Applied Physics Letters\/} {\bf 90} 043113 ISSN 0003-6951,
  1077-3118

\bibitem{PerisanulinearDuffingregimes2010}
Perisanu S, Barois T, Ayari A, Poncharal P, Choueib M, Purcell S~T and Vincent
  P 2010 {\em Physical Review B\/} {\bf 81} 165440

\bibitem{PigeauObservationphononicMollow2015}
Pigeau B, Rohr S, {Mercier de L\'epinay} L, Gloppe A, Jacques V and Arcizet O
  2015 {\em Nature Communications\/} {\bf 6} ISSN 2041-1723

\bibitem{NicholDisplacementdetectionsilicon2008}
Nichol J~M, Hemesath E~R, Lauhon L~J and Budakian R 2008 {\em Applied Physics
  Letters\/} {\bf 93} 193110 ISSN 0003-6951, 1077-3118

\bibitem{NicholControllingnonlinearitysilicon2009}
Nichol J~M, Hemesath E~R, Lauhon L~J and Budakian R 2009 {\em Applied Physics
  Letters\/} {\bf 95} 123116 ISSN 0003-6951

\bibitem{Gil-SantosNanomechanicalmasssensing2010}
{Gil-Santos} E, Ramos D, Mart\'inez J, {Fern\'andez-Reg\'ulez} M, Garc\'ia R,
  San~Paulo A, Calleja M and Tamayo J 2010 {\em Nature Nanotechnology\/} {\bf
  5} 641--645 ISSN 1748-3387

\bibitem{Gil-SantosOpticalbackactionsilicon2013}
{Gil-Santos} E, Ramos D, Pini V, Llorens J, {Fern\'andez-Reg\'ulez} M, Calleja
  M, Tamayo J and Paulo A~S 2013 {\em New Journal of Physics\/} {\bf 15} 035001
  ISSN 1367-2630

\bibitem{RamosOptomechanicsSiliconNanowires2012}
Ramos D, {Gil-Santos} E, Pini V, Llorens J~M, {Fern\'andez-Reg\'ulez} M,
  San~Paulo A, Calleja M and Tamayo J 2012 {\em Nano Letters\/} {\bf 12}
  932--937 ISSN 1530-6984

\bibitem{RamosSiliconnanowireswhere2013}
Ramos D, {Gil-Santos} E, Malvar O, Llorens J~M, Pini V, Paulo A~S, Calleja M
  and Tamayo J 2013 {\em Scientific Reports\/} {\bf 3} ISSN 2045-2322

\bibitem{BelovMechanicalresonanceclamped2008}
Belov M, Quitoriano N~J, Sharma S, Hiebert W~K, Kamins T~I and Evoy S 2008 {\em
  Journal of Applied Physics\/} {\bf 103} 074304 ISSN 0021-8979, 1089-7550

\bibitem{FengVeryHighFrequency2007}
Feng X~L, He R, Yang P and Roukes M~L 2007 {\em Nano Letters\/} {\bf 7}
  1953--1959 ISSN 1530-6984

\bibitem{SiriaElectronbeamdetection2017}
Siria A and Nigu\`es A 2017 {\em Scientific Reports\/} {\bf 7} 11595 ISSN
  2045-2322

\bibitem{TavernarakisOptomechanicshybridcarbon2018}
Tavernarakis A, Stavrinadis A, Nowak A, Tsioutsios I, Bachtold A and Verlot P
  2018 {\em Nature Communications\/} {\bf 9} 662 ISSN 2041-1723

\bibitem{TsioutsiosRealTimeMeasurementNanotube2017}
Tsioutsios I, Tavernarakis A, Osmond J, Verlot P and Bachtold A 2017 {\em Nano
  Letters\/} {\bf 17} 1748--1755 ISSN 1530-6984

\bibitem{ClelandFoundationsNanomechanicsSolidState2003}
Cleland A~N 2003 {\em Foundations of {{Nanomechanics From Solid}}-{{State
  Theory}} to {{Device Applications}}\/} (Berlin, Heidelberg: {Springer Berlin
  Heidelberg}) ISBN 978-3-662-05287-7 3-662-05287-3

\bibitem{PootMechanicalsystemsquantum2012}
Poot M and {van der Zant} H~S~J 2012 {\em Physics Reports\/} {\bf 511} 273--335
  ISSN 0370-1573

\bibitem{ReifFundamentalsstatisticalthermal1965}
Reif F 1965 {\em Fundamentals of Statistical and Thermal Physics\/} McGraw-Hill
  series in fundamentals of physics (New York [etc.]: {McGraw-Hill}) ISBN
  978-0-07-051800-1

\bibitem{Kubofluctuationdissipationtheorem1966}
Kubo R 1966 {\em Reports on Progress in Physics\/} {\bf 29} 255 ISSN 0034-4885

\bibitem{RobinsPhasenoisesignal1982}
Robins W~P 1982 {\em Phase Noise in Signal Sources : (Theory and
  Applications)\/} IEE telecommunications series (London: {Peregrinus}) ISBN
  978-0-906048-76-4

\bibitem{SansaFrequencyfluctuationssilicon2016}
Sansa M, Sage E, Bullard E~C, G\'ely M, Alava T, Colinet E, Naik A~K,
  Villanueva L~G, Duraffourg L, Roukes M~L, Jourdan G and Hentz S 2016 {\em Nat
  Nano\/} {\bf 11} 552--558 ISSN 1748-3387

\bibitem{MoserUltrasensitiveforcedetection2013}
Moser J, G\"uttinger J, Eichler A, Esplandiu M~J, Liu D~E, Dykman M~I and
  Bachtold A 2013 {\em Nature Nanotechnology\/} {\bf 8} 493--496 ISSN 1748-3387

\bibitem{FaustNonadiabaticDynamicsTwo2012}
Faust T, Rieger J, Seitner M~J, Krenn P, Kotthaus J~P and Weig E~M 2012 {\em
  Physical Review Letters\/} {\bf 109} 037205

\bibitem{FaustCoherentcontrolclassical2013}
Faust T, Rieger J, Seitner M~J, Kotthaus J~P and Weig E~M 2013 {\em Nature
  Physics\/} {\bf 9} 485--488 ISSN 1745-2473

\bibitem{SpreeuwClassicalrealizationstrongly1990}
Spreeuw R~J~C, {van Druten} N~J, Beijersbergen M~W, Eliel E~R and Woerdman J~P
  1990 {\em Physical Review Letters\/} {\bf 65} 2642--2645

\bibitem{OkamotoCoherentphononmanipulation2013}
Okamoto H, Gourgout A, Chang C~Y, Onomitsu K, Mahboob I, Chang E~Y and
  Yamaguchi H 2013 {\em Nature Physics\/} {\bf 9} 480--484 ISSN 1745-2473

\bibitem{DegenQuantumsensing2017}
Degen C~L, Reinhard F and Cappellaro P 2017 {\em Reviews of Modern Physics\/}
  {\bf 89} 035002

\bibitem{BraakmanNonlinearmotionmechanical2014}
Braakman F~R, Cadeddu D, T\"ut\"unc\"uoglu G, Matteini F, R\"uffer D, i~Morral
  A~F and Poggio M 2014 {\em Applied Physics Letters\/} {\bf 105} 173111 ISSN
  0003-6951, 1077-3118

\bibitem{Eichlerparametricsymmetrybreaking2018}
Eichler A, Heugel T~L, Leuch A, Degen C~L, Chitra R and Zilberberg O 2018 {\em
  Applied Physics Letters\/} {\bf 112} 233105 ISSN 0003-6951

\bibitem{SilvaNonlinearFlexuralFlexuralTorsionalDynamics1978}
da~Silva M~R~M~C and Glynn C~C 1978 {\em Journal of Structural Mechanics\/}
  {\bf 6} 449--461 ISSN 0360-1218

\bibitem{SilvaNonlinearFlexuralFlexuralTorsionalDynamics1978a}
da~Silva M~R~M~C and Glynn C~C 1978 {\em Journal of Structural Mechanics\/}
  {\bf 6} 437--448 ISSN 0360-1218

\bibitem{SchmidFundamentalsNanomechanicalResonators2016}
Schmid S, Villanueva L~G and Roukes M~L 2016 {\em Fundamentals of
  {{Nanomechanical Resonators}}\/} ({Springer International Publishing}) ISBN
  978-3-319-28689-1

\bibitem{CadedduNanomechanicsscanningprobe}
Cadeddu D {\em Nanomechanics and Scanning Probe Microscopy with Nanowires\/}
  Ph.D. thesis

\bibitem{LifshitzNonlineardynamicsnanomechanical}
Lifshitz R and Cross M~C {\em Review of nonlinear dynamics and complexity\/}
  {\bf 1} 1--52

\bibitem{EichlerNonlineardampingmechanical2011}
Eichler A, Moser J, Chaste J, Zdrojek M, {Wilson-Rae} I and Bachtold A 2011
  {\em Nature Nanotechnology\/} {\bf 6} 339--342 ISSN 1748-3395

\bibitem{RugarAtomicForceMicroscopy2008}
Rugar D and Hansma P 2008 {\em Physics Today\/} {\bf 43} 23 ISSN 0031-9228

\bibitem{AkamineImprovedatomicforce1990}
Akamine S, Barrett R~C and Quate C~F 1990 {\em Applied Physics Letters\/} {\bf
  57} 316--318 ISSN 0003-6951

\bibitem{GonzalezBrownianmotionmass1994}
Gonz\'alez G~I and Saulson P~R 1994 {\em The Journal of the Acoustical Society
  of America\/} {\bf 96} 207--212 ISSN 0001-4966

\bibitem{CagnoliDampingdilutionfactor2000}
Cagnoli G, Hough J, DeBra D, Fejer M~M, Gustafson E, Rowan S and Mitrofanov V
  2000 {\em Physics Letters A\/} {\bf 272} 39--45 ISSN 0375-9601

\bibitem{TsaturyanUltracoherentnanomechanicalresonators2017}
Tsaturyan Y, Barg A, Polzik E~S and Schliesser A 2017 {\em Nature
  Nanotechnology\/} {\bf advance online publication} ISSN 1748-3387

\bibitem{GhadimiElasticstrainengineering2018}
Ghadimi A~H, Fedorov S~A, Engelsen N~J, Bereyhi M~J, Schilling R, Wilson D~J
  and Kippenberg T~J 2018 {\em Science\/} {\bf 360} 764--768 ISSN 0036-8075,
  1095-9203

\bibitem{MarshallQuantumSuperpositionsMirror2003}
Marshall W, Simon C, Penrose R and Bouwmeester D 2003 {\em Physical Review
  Letters\/} {\bf 91} 130401

\bibitem{RugarAdventuresattonewtonforce2001}
Rugar D, Stipe B~C, Mamin H~J, Yannoni C~S, Stowe T~D, Yasumura K~Y and Kenny
  T~W 2001 {\em Applied Physics A\/} {\bf 72} S3--S10 ISSN 0947-8396, 1432-0630

\bibitem{StipeMagneticDissipationFluctuations2001}
Stipe B~C, Mamin H~J, Stowe T~D, Kenny T~W and Rugar D 2001 {\em Physical
  Review Letters\/} {\bf 86} 2874--2877 00111

\bibitem{TaoPermanentreductiondissipation2015}
Tao Y, Navaretti P, Hauert R, Grob U, Poggio M and Degen C~L 2015 {\em
  Nanotechnology\/} {\bf 26} 465501 ISSN 0957-4484

\bibitem{StipeNoncontactFrictionForce2001}
Stipe B~C, Mamin H~J, Stowe T~D, Kenny T~W and Rugar D 2001 {\em Physical
  Review Letters\/} {\bf 87} 096801

\bibitem{KuehnDielectricFluctuationsOrigins2006}
Kuehn S, Loring R~F and Marohn J~A 2006 {\em Physical Review Letters\/} {\bf
  96} 156103

\bibitem{Sazonovatunablecarbonnanotube2004}
Sazonova V, Yaish Y, \"Ust\"unel H, Roundy D, Arias T~A and McEuen P~L 2004
  {\em Nature\/} {\bf 431} 284--287 ISSN 0028-0836

\bibitem{BunchElectromechanicalResonatorsGraphene2007}
Bunch J~S, van~der Zande A~M, Verbridge S~S, Frank I~W, Tanenbaum D~M, Parpia
  J~M, Craighead H~G and McEuen P~L 2007 {\em Science\/} {\bf 315} 490--493
  ISSN 0036-8075, 1095-9203

\bibitem{MoserNanotubemechanicalresonators2014}
Moser J, Eichler A, G\"uttinger J, Dykman M~I and Bachtold A 2014 {\em Nature
  Nanotechnology\/} {\bf advance online publication} ISSN 1748-3387

\bibitem{GysinLowtemperatureultrahigh2011}
Gysin U, Rast S, Kisiel M, Werle C and Meyer E 2011 {\em Review of Scientific
  Instruments\/} {\bf 82} 023705 ISSN 0034-6748, 1089-7623

\bibitem{SekaricNanomechanicalresonantstructures2002}
Sekaric L, Carr D~W, Evoy S, Parpia J~M and Craighead H~G 2002 {\em Sensors and
  Actuators A: Physical\/} {\bf 101} 215--219 ISSN 0924-4247

\bibitem{AzakNanomechanicaldisplacementdetection2007}
Azak N~O, Shagam M~Y, Karabacak D~M, Ekinci K~L, Kim D~H and Jang D~Y 2007 {\em
  Applied Physics Letters\/} {\bf 91} 093112 ISSN 0003-6951

\bibitem{LiBottomupassemblylargearea2008}
Li M, Bhiladvala R~B, Morrow T~J, Sioss J~A, Lew K~K, Redwing J~M, Keating C~D
  and Mayer T~S 2008 {\em Nature Nanotechnology\/} {\bf 3} 88--92 ISSN
  1748-3387

\bibitem{FaveroFluctuatingnanomechanicalsystem2009}
Favero I, Stapfner S, Hunger D, Paulitschke P, Reichel J, Lorenz H, Weig E~M
  and Karrai K 2009 {\em Optics Express\/} {\bf 17} 12813 ISSN 1094-4087

\bibitem{TannerHighQGaNnanowire2007}
Tanner S~M, Gray J~M, Rogers C~T, Bertness K~A and Sanford N~A 2007 {\em
  Applied Physics Letters\/} {\bf 91} 203117 ISSN 0003-6951

\bibitem{HeSelfTransducingSiliconNanowire2008}
He R, Feng X~L, Roukes M~L and Yang P 2008 {\em Nano Letters\/} {\bf 8}
  1756--1761 ISSN 1530-6984

\bibitem{TruittEfficientSensitiveCapacitive2007}
Truitt P~A, Hertzberg J~B, Huang C~C, Ekinci K~L and Schwab K~C 2007 {\em Nano
  Letters\/} {\bf 7} 120--126 ISSN 1530-6984

\bibitem{SaniiHighSensitivityDeflection2010}
Sanii B and Ashby P~D 2010 {\em Physical Review Letters\/} {\bf 104} 147203

\bibitem{FuDeterminingdirectionnanowire2017}
Fu C, Deng W, Zou L, Zhu W, Wang N and Xue F 2017 {\em arXiv:1711.04446
  [cond-mat]\/} (\textit{Preprint} \eprint{1711.04446})

\bibitem{WeymouthNoncontactlateralforce2017}
Weymouth A~J 2017 {\em Journal of Physics: Condensed Matter\/} {\bf 29} 323001
  ISSN 0953-8984

\bibitem{PfeifferLateralforcemeasurementsdynamic2002}
Pfeiffer O, Bennewitz R, Baratoff A, Meyer E and Gr\"utter P 2002 {\em Physical
  Review B\/} {\bf 65} 161403

\bibitem{GiessiblFrictiontracedsingle2002}
Giessibl F~J, Herz M and Mannhart J 2002 {\em Proceedings of the National
  Academy of Sciences\/} {\bf 99} 12006--12010 ISSN 0027-8424, 1091-6490

\bibitem{KawaiDynamiclateralforce2005}
Kawai S, Kitamura S~i, Kobayashi D and Kawakatsu H 2005 {\em Applied Physics
  Letters\/} {\bf 87} 173105 ISSN 0003-6951, 1077-3118

\bibitem{KawaiDirectmappinglateral2009}
Kawai S, Sasaki N and Kawakatsu H 2009 {\em Physical Review B\/} {\bf 79}
  195412

\bibitem{KawaiUltrasensitivedetectionlateral2010}
Kawai S, Glatzel T, Koch S, Such B, Baratoff A and Meyer E 2010 {\em Physical
  Review B\/} {\bf 81} 085420

\bibitem{VilhenaSlipperyeverydirection2018}
Vilhena J~G and P\'erez R 2018 {\em Nature Materials\/} {\bf 17} 852 ISSN
  1476-4660

\bibitem{SchmidExchangeBiasDomain2010}
Schmid I, Marioni M~A, Kappenberger P, Romer S, {Parlinska-Wojtan} M, Hug H~J,
  Hellwig O, Carey M~J and Fullerton E~E 2010 {\em Physical Review Letters\/}
  {\bf 105} 197201

\bibitem{DegenNanoscalemagneticresonance2009}
Degen C~L, Poggio M, Mamin H~J, Rettner C~T and Rugar D 2009 {\em Proceedings
  of the National Academy of Sciences\/} {\bf 106} 1313--1317 ISSN 0027-8424,
  1091-6490

\bibitem{KirtleyFundamentalstudiessuperconductors2010}
Kirtley J~R 2010 {\em Reports on Progress in Physics\/} {\bf 73} 126501 ISSN
  0034-4885

\bibitem{KaidatzisTorsionalresonancemode2013}
Kaidatzis A and {Garc\'ia-Mart\'in} J~M 2013 {\em Nanotechnology\/} {\bf 24}
  165704 ISSN 0957-4484

\bibitem{DengMetalcoatedcarbonnanotube2004}
Deng Z, Yenilmez E, Leu J, Hoffman J~E, Straver E~W~J, Dai H and Moler K~A 2004
  {\em Applied Physics Letters\/} {\bf 85} 6263--6265 ISSN 0003-6951

\bibitem{PhillipsHighresolutionmagnetic2002}
Phillips G~N, Siekman M, Abelmann L and Lodder J~C 2002 {\em Applied Physics
  Letters\/} {\bf 81} 865--867 ISSN 0003-6951

\bibitem{LauPropertiesapplicationscobaltbased2002}
Lau Y~M, Chee P~C, Thong J~T~L and Ng V 2002 {\em Journal of Vacuum Science \&
  Technology A: Vacuum, Surfaces, and Films\/} {\bf 20} 1295--1302 ISSN
  0734-2101

\bibitem{KoblischkaImprovementslateralresolution2003}
Koblischka M~R, Hartmann U and Sulzbach T 2003 {\em Thin Solid Films\/} {\bf
  428} 93--97 ISSN 0040-6090

\bibitem{HubmannEpitaxialGrowthRoomTemperature2016}
Hubmann J, Bauer B, K\"orner H~S, Furthmeier S, Buchner M, Bayreuther G,
  Dirnberger F, Schuh D, Back C~H, Zweck J, Reiger E and Bougeard D 2016 {\em
  Nano Letters\/} {\bf 16} 900--905 ISSN 1530-6984

\bibitem{RosselActivemicroleversminiature1996}
Rossel C, Bauer P, Zech D, Hofer J, Willemin M and Keller H 1996 {\em Journal
  of Applied Physics\/} {\bf 79} 8166--8173 ISSN 0021-8979, 1089-7550

\bibitem{HarrisIntegratedmicromechanicalcantilever1999}
Harris J~G~E, Awschalom D~D, Matsukura F, Ohno H, Maranowski K~D and Gossard
  A~C 1999 {\em Applied Physics Letters\/} {\bf 75} 1140--1142 ISSN 0003-6951,
  1077-3118

\bibitem{Schendelmethodcalibrationmagnetic2000}
van Schendel P~J~A, Hug H~J, Stiefel B, Martin S and G\"untherodt H~J 2000 {\em
  Journal of Applied Physics\/} {\bf 88} 435--445 ISSN 0021-8979, 1089-7550

\bibitem{Ciobanu3DMRmicroscopy2002}
Ciobanu L, Seeber D~A and Pennington C~H 2002 {\em Journal of Magnetic
  Resonance\/} {\bf 158} 178--182 ISSN 1090-7807

\bibitem{PoggioForcedetectednuclearmagnetic2010}
Poggio M and Degen C~L 2010 {\em Nanotechnology\/} {\bf 21} 342001 ISSN
  0957-4484 00041

\bibitem{PoggioForceDetectedNuclearMagnetic2018}
Poggio M and Herzog B~E 2018 Force-{{Detected Nuclear Magnetic Resonance}} {\em
  Micro and {{Nano Scale NMR}}\/} ({Wiley-Blackwell}) pp 381--420 ISBN
  978-3-527-69728-1

\bibitem{herzog_boundary_2014}
Herzog B~E, Cadeddu D, Xue F, Peddibhotla P and Poggio M 2014 {\em Applied
  Physics Letters\/} {\bf 105} 043112 ISSN 0003-6951, 1077-3118

\bibitem{NicholNanoscalemagneticresonance2014}
Nichol J 2014 {\em Nanoscale Magnetic Resonance Imaging Using Silicon Nanowire
  Oscillators\/} Ph.D. thesis University of Illinois at Urbana-Champaign 00000

\bibitem{KisielNoncontactAtomicForce2015}
Kisiel M, Pellegrini F, Santoro G~E, Samadashvili M, Pawlak R, Benassi A, Gysin
  U, Buzio R, Gerbi A, Meyer E and Tosatti E 2015 {\em Physical Review
  Letters\/} {\bf 115} 046101

\bibitem{VolokitinQuantumFriction2011}
Volokitin A~I and Persson B~N~J 2011 {\em Physical Review Letters\/} {\bf 106}
  094502

\bibitem{HalbertalNanoscalethermalimaging2016}
Halbertal D, Cuppens J, Shalom M~B, Embon L, Shadmi N, Anahory Y, Naren H~R,
  Sarkar J, Uri A, Ronen Y, Myasoedov Y, Levitov L~S, Joselevich E, Geim A~K
  and Zeldov E 2016 {\em Nature\/} {\bf 539} 407--410 ISSN 0028-0836

\end{thebibliography}
\end{document}